%% file: final-arxiv.tex
\documentclass[article]{llncs}

\usepackage{pslatex}
\usepackage{amsmath}
\usepackage{amssymb}
\usepackage{leftidx}
\usepackage{epsfig}
\usepackage{paralist}
\usepackage{graphics}
\usepackage{stmaryrd}
\usepackage{txfonts}
\usepackage{framed}
\usepackage{makecell}
\usepackage{subfig}
\usepackage{wrapfig}
\usepackage[bookmarks,unicode,colorlinks=true]{hyperref}
\usepackage{csquotes}

\usepackage[version=0.96]{pgf}
\usepackage{tikz}
\usetikzlibrary{arrows,shapes,snakes,automata,backgrounds,petri,positioning}
\usepackage[latin1]{inputenc}

\usepackage{graphicx}

\input commands

\usepackage{thm-restate} 

\usepackage{setspace}

\newif\ifLongVersion\LongVersiontrue

\begin{document}

\setlength{\belowdisplayskip}{2pt} \setlength{\belowdisplayshortskip}{1pt}
\setlength{\abovedisplayskip}{2pt} \setlength{\abovedisplayshortskip}{1pt}

\title{Structural Invariants for the Verification of Systems with
  Parameterized Architectures}

\ifLongVersion
\author{Marius Bozga\inst{1}, Javier Esparza\inst{2}, Radu Iosif\inst{1}, Joseph Sifakis\inst{1}\\
and Christoph Welzel\inst{2}}
\institute{Univ. Grenoble Alpes, CNRS, Grenoble INP\footnote{Institute of Engineering Univ. Grenoble Alpes}, Verimag
   \and
   Technische Universit\"at M\"unchen
}
\else
\author{~}
\institute{~}
\fi

\maketitle
\input abstract
\input body
\input conclusions
\input appendix

\bibliographystyle{splncs03} \bibliography{refs}

\end{document}

%% file: commands.tex
\newtheorem{fact}{Fact}

\renewcommand{\mod}{~\mathrm{mod}~}

\newcommand{\set}[1]{\{ #1 \}}
\newcommand{\tuple}[1]{\langle #1 \rangle}
\renewcommand{\vec}[1]{\mathbf #1}

\newcommand{\isdef}{\stackrel{\scriptscriptstyle{\mathsf{def}}}{=}}

\newcommand{\card}[1]{{||{#1}||}}

\newcommand{\arrow}[2]{\xrightarrow{{\scriptscriptstyle #1}}_{{\scriptscriptstyle #2}}}

\newcommand{\nat}{{\bf \mathbb{N}}}

\renewcommand{\paragraph}[1]{\noindent{\bf #1}}

\newcommand{\I}{\mathcal{I}}
\newcommand{\Universe}{\mathfrak{U}}
\newcommand{\true}{\top}

\newcommand{\vars}{\mathsf{Var}}
\newcommand{\Vars}{\mathsf{SVar}}

\newcommand{\preds}{\mathsf{Pred}}

\newcommand{\anet}{\mathsf{N}}
\newcommand{\amarkednet}{\mathcal{N}}
\newcommand{\places}{S}
\newcommand{\placeset}{G}
\newcommand{\trans}{T}
\newcommand{\atrans}{\mathfrak{t}}
\newcommand{\edges}{E}
\newcommand{\pre}[1]{\leftidx{^\bullet}{\text{${#1}$}}}
\newcommand{\post}[1]{{#1}^\bullet}

\newcommand{\amark}{\mathrm{m}}
\newcommand{\reach}[1]{\mathcal{R}({#1})}

\newcommand{\acomptype}{\mathcal{C}}

\newcommand{\typeno}[2]{{#1}^{\scriptscriptstyle{{#2}}}}
\newcommand{\typeof}[1]{\mathit{type}({#1})}
\newcommand{\ports}{\mathsf{P}}
\newcommand{\states}{\mathsf{S}}
\newcommand{\initstate}{{s_0}}
\newcommand{\rules}{\Delta}
\newcommand{\asys}{\mathcal{S}}
\newcommand{\interform}{\Gamma}

\newcommand{\trapconstraint}[1]{\Theta({#1})}

\newcommand{\statevar}[1]{X_{#1}}

\newcommand{\il}[1]{$\mathsf{IL}{#1}$}
\newcommand{\ilmodels}{\models_{\scriptscriptstyle{\mathsf{IL}}}}

\newcommand{\wss}[1]{$\mathsf{WS}\kappa\mathsf{S}$}
\newcommand{\wsones}{$\mathsf{WS}1\mathsf{S}$}
\newcommand{\wsmodels}{\models_{\scriptscriptstyle{\mathsf{WS}\kappa\mathsf{S}}}}

\newcommand{\mso}{$\mathsf{MSO}$}

\newcommand{\wssroot}{\overline{\epsilon}}

\newcommand{\strord}{\sqsubseteq}

\newcommand{\flow}{\mathfrak{F}}
\DeclareMathOperator{\aclause}{\mathfrak{C}}
\DeclareMathOperator{\intersectspre}{\mathit{intersects-pre}_{\asys}^{\aclause}}
\DeclareMathOperator{\intersectspost}{\mathit{intersects-post}_{\asys}^{\aclause}}
\DeclareMathOperator{\trappred}{\mathit{trap-pred}_{\asys}}
\DeclareMathOperator{\marking}{\mathit{marking}_{\asys}}
\DeclareMathOperator{\intersection}{\mathit{intersection}_{\asys}}
\DeclareMathOperator{\initially}{\mathit{initially-marked}_{\asys}}
\DeclareMathOperator{\uniqueinitially}{\mathit{unique-init}_{\asys}}
\DeclareMathOperator{\uniqueintersection}{\mathit{unique-intersection}_{\asys}}
\DeclareMathOperator{\trapconstraintpred}{\mathit{trap-invariant}_{\asys}}
\DeclareMathOperator{\flowinvariant}{\mathit{1-invariant}_{\asys}}
\DeclareMathOperator{\flowpred}{\mathit{1-pred}_{\asys}}
\DeclareMathOperator{\uniquepre}{\mathit{uniquepre}_{\asys}^{\aclause}}
\DeclareMathOperator{\uniquepreex}{\mathit{unique-pre-ex}_{\asys}^{\aclause}}
\DeclareMathOperator{\disjointprebroadcast}{\mathit{disjoint-pre-broadcast}_{\asys}^{\aclause}}
\DeclareMathOperator{\disjointpreex}{\mathit{disjoint-pre-ex}_{\asys}^{\aclause}}
\DeclareMathOperator{\uniqueprebroadcast}{\mathit{unique-pre-broadcast}_{\asys}^{\aclause}}
\DeclareMathOperator{\uniquepost}{\mathit{uniquepost}_{\asys}^{\aclause}}

\renewcommand{\succ}{\mathrm{succ}}
\newcommand{\apred}{\mathsf{pr}}

\renewcommand{\proof}[1]{\ifLongVersion \noindent\emph{Proof}: {#1} \vspace*{\baselineskip}\else\fi}

%% file: abstract.tex
We consider parameterized concurrent systems consisting of a finite
but unknown number of components, obtained by replicating a given set
of finite state automata.  Components communicate by executing atomic
interactions whose participants update their states simultaneously. We
introduce an interaction logic to specify both the type of
interactions (e.g.\ rendez-vous, broadcast) and the topology of the
system (e.g.\ pipeline, ring). The logic can be easily embedded in
monadic second order logic of $\kappa\geq1$ successors (\wss{\kappa}),
and is therefore decidable.

Proving safety properties of such a parameterized system, like
deadlock freedom or mutual exclusion, requires to infer an
inductive invariant that contains all reachable states of all system instances, and
no unsafe state. We present a method to automatically synthesize 
inductive invariants directly from the formula describing the interactions,
without costly fixed point iterations. We experimentally prove that
this invariant is strong enough to verify safety properties of a large number
of systems including textbook examples (dining philosophers,
synchronization schemes), classical mutual exclusion algorithms, 
cache-coherence protocols and self-stabilization algorithms, for an arbitrary number of components.

%% file: body.tex
\section{Introduction}

The problem of parameterized verification asks whether a system composed
of $n$ replicated processes is safe, for all $n \geq 2$. By safety we
mean that every execution of the system stays clear of a set of global
error configurations, such as deadlocks or mutual exclusion
violations. Even if we assume each process to be finite-state and
every interaction to be a synchronization of actions without exchange
of data, ranging over large or infinite domains, the problem remains
challenging because we ask for a general proof of safety, that works
for any number of processes.

In general, parameterized verification is undecidable, even if processes only manipulate data
from a bounded  domain \cite{AptKozen86}. Various restrictions of communication
and architecture\footnote{We use the term architecture for the shape
  of the graph along which the interactions take place.} define decidable subproblems
\cite{ClarkeGrumbergBrowne86,GermanSistla92,EmersonNamjoshi95,AminofKotekRubinSpegniVeith17}.
Seminal works consider \emph{rendez-vous} communication, with
 participants placed in a ring \cite{ClarkeGrumbergBrowne86,EmersonNamjoshi95} or a clique
\cite{GermanSistla92} of arbitrary size. Recently, \mso-definable graphs (with bounded
tree- and clique-width) and point-to-point rendez-vous communication
have been considered \cite{AminofKotekRubinSpegniVeith17}. Most
approaches to define decidable problems focus on manually proving a
\emph{cut-off} bound $c\geq2$ such that correctness for at most $c$ processes 
implies correctness for any number of processes
\cite{ClarkeGrumbergBrowne86,EmersonNamjoshi95,EmersonK00,AusserlechnerJK16,JacobsS18}. Other methods
identify systems with well-structured transition relations \cite{GermanSistla92,AbdullaCJT96,FinkelS01}.
An exhaustive chart of decidability results for verification of parameterized
systems is drawn in \cite{BloemJacobsKhalimovKonnovRubinVeithWidder15}. When decidability
is not of concern, over-approximating and semi-algorithmic techniques
such as \emph{regular model checking}
\cite{KestenMalerMarcusPnueliShahar01,AbdullaHendaDelzannoRezine07},
SMT-based \emph{bounded model checking}
\cite{AlbertiGhilardiSharygina14,ConchonGoelKrsticMebsoutZaidi12},
\emph{abstraction}
\cite{BaukusBensalemLakhnechStahl00,BouajjaniHabermehlVojnar04} and
\emph{automata learning} \cite{ChenHongLinRummer17} can be used to
deal with more general classes of systems.

The efficiency of a verification method crucially relies on its
ability to synthesize an \emph{inductive safety invariant}, i.e.,
an infinite set of configurations that contains the initial
configurations, is closed under the transition relation, and excludes
the error configurations. In general, automatically synthesizing
invariants requires computationally expensive fixpoint iterations
\cite{CousotCousot79}. In the particular case of parameterized systems,
invariants can be either \emph{global}, relating the local states of
all processes \cite{DamsLakhnechSteffen01}, or \emph{modular},
relating the local states of a few processes whose identity is irrelevant
\cite{PnueliRuahZuck01,ClarkeTalupurVeith06}.

\paragraph{\em Our Contributions.}
The novelty of the approach described in this paper is three-fold: 
\begin{compactenum}
\item The architecture of the system is not fixed a~priori, but given
  as a parameter of the verification problem. In fact, we describe
  parameterized systems using the Behavior-Inter\-action-Priorities (BIP)
  framework \cite{BasuBBCJNS11}, in which processes are instances of
  finite-state \emph{component types}, whose interfaces are sets of
  \emph{ports}, labeling transitions between local states, and
  interactions are sets of strongly synchronizing ports, described by
  formulae of an \emph{interaction logic}. An interaction formula
  captures the architecture of the interactions (pipeline, ring,
  clique, tree) and the communication scheme (rendez-vous, broadcast),
  which are not hardcoded, but rather specified by the 
  designer of the system. 
\item We synthesize parameterized invariants directly from the
  interaction formula of a system, without iterating its transition
  relation. Such invariants depend only on the structure (and not on
  the operational semantics) of an infinite family of Petri Nets, one
  for each instance of the system, being thus \emph{structural}
  invariants. Essentially, the invariants we infer use the
  \emph{traps}\footnote{Called in this way by analogy with the notion
    of traps for Petri Nets \cite{Sifakis78}.} of the system, which
  are sets $W$ of local states with the property that, if a process is
  initially in a state from $W$, then always some process will be in a
  state from $W$. Following \cite{BensalemBNS09,BozgaIosifSifakis19a},
  we call them (parameterized) \emph{trap invariants}.  Computing trap
  invariants only requires a simple syntactic transformation of the
  interaction formula and the result is expressed using \wss{\kappa},
  the weak monadic second order logic of $\kappa\geq1$ successor
  functions. Thus invariant computation is very cheap, and the
  verification problem (proving the emptiness of the intersection
  between the invariant and the set of error states) is reduced to the
  unsatisfiability of a \wss{\kappa} formula with a single quantifier
  alternation. In practice, this check can be carried out quite
  efficiently by existing tools, such as \textsc{Mona} \cite{mona}.
\item We refine the approach by considering another type of
  invariants, called \emph{$1$-invariants}, that can also be derived
  cheaply from the interaction formula of the system. We show that
  $1$-invariants in conjunction with trap invariants successfully
  verify additional examples.
\end{compactenum}

\paragraph{Comparison to related work.}
Trap invariants have been very successfully used in the verification
of non-parameterized systems \cite{BensalemBNS09,ELMMN14,BFHH16}.  The
technique was lifted to parameterized systems in
\cite{BozgaIosifSifakis19a}, but the work there is only applicable to
clique architectures, in which processes are indistinguishable, and
the system can be described by one single Petri Net with an infinite
family of initial markings. Here, for the first time, we show that the
trap technique can be extended to pipelines, token rings and trees,
where the system is defined by an infinite family of Petri Nets, each
with a different structure. These systems cannot be analyzed using the
techniques of \cite{GermanSistla92,AbdullaCJT96,FinkelS01}, because
they do not yield well-structured transition systems. Contrary to
\cite{ClarkeGrumbergBrowne86,EmersonNamjoshi95,EmersonK00,AusserlechnerJK16,JacobsS18},
our approach does not require a manual cut-off proof. Contrary to
regular model checking and automata learning
\cite{AbdullaHendaDelzannoRezine07,ChenHongLinRummer17}, it does not
require any symbolic state-space exploration. Finally, our approach
produces an explanation of why the property holds in terms of the trap
invariant and $1$-invariants used. Summarizing, our approach provides
a comparatively cheap technique for parameterized verification, that
succeeds in numerous cases. It is ideal as preprocessing step, that
can very quickly lead to success with a very clear explanation of why
the property holds, and otherwise provides at least a strong invariant
that can be used for further analysis.

\begin{figure}[ht]
  \vspace*{-\baselineskip}
  \begin{center}
    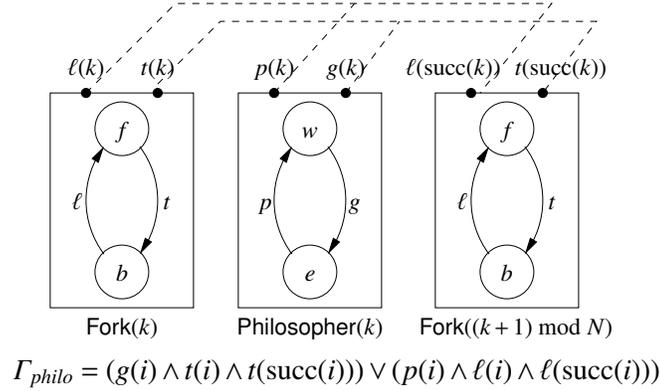
  \end{center}
  \vspace*{-\baselineskip}
  \caption{Parameterized Dining Philosophers}
  \label{fig:philosophers}
  \vspace*{-\baselineskip}
  \end{figure}

\paragraph{\em Running Example.}
Consider the dining philosophers system in
Fig. \ref{fig:philosophers}, 
consisting of $n\geq2$ components of type
$\mathsf{Fork}$ and $\mathsf{Philosopher}$ respectively, placed in a
ring of size $2n$. The $k$-th philosopher has a left fork, of index
$k$, and a right fork, of index $(k+1) \mod n$. Each component is an
instance of a finite state automaton with states $f$(ree) and $b$(usy)
for $\mathsf{Fork}$, respectively $w$(aiting) and $e$(ating) for
$\mathsf{Philosopher}$. A fork goes from state $f$ to $b$ via a
$t$(ake) transition and from $f$ to $b$ via a $\ell$(eave)
transition. A philosopher goes from $w$ to $e$ via a $g$(et)
transition and from $e$ to $w$ via a $p$(ut) transition. The $g$
action of the $k$-th philosopher is executed jointly with the $t$
actions of the $k$-th and $[(k+1) \mod n]$-th forks, in other words, the
philosopher takes both its left and right forks
simultaneously. Similarly, the $p$ action of the $k$-th philosopher is
executed simultaneously with the $\ell$ action of the $k$-th and
$[(k+1) \mod n]$-th forks, i.e.\ each philosopher leaves both its left
and right forks at the same time. We describe these interactions by
the \emph{interaction formula}: 
\begin{equation}
\interform_{\mathit{philo}} = \;\; (g(i) \wedge t(i) \wedge
  t(\succ(i))) \;\; \vee  \;\; (p(i) \wedge \ell(i) \wedge \ell(\succ(i))) \enspace.
\label{interform}
\end{equation}
where the free variable $i$ refers at some arbitrary component index.

Intuitively, the transitions of the system with $n$ dining
philosophers and $n$ forks are given by the \emph{minimal models} of
the disjuncts of $\interform_{\mathit{philo}}$ with universe $\{0,1,
\ldots, n-1\}$, and $\succ$ interpreted as ``successor modulo $n$'.
In particular, for each $0 \leq k \leq n-1$ the first disjunct has a
minimal model that interprets the predicates $g$ and $t$ as the sets
$\set{k}$ and $\set{k,(k+1)\mod n}$. This model describes the
interaction in which the $k$-th philosopher takes a $g$-transition
(from waiting to eating), while, simultaneously, the $k$-th and
$(k+1)$-th forks take $t$-transitions (from free to busy). This is
graphically represented by one of the dashed lines in
Fig. \ref{fig:philosophers}. Observe that the ring topology of the
system is implicit in the modulo-$n$ interpretation of the successor
function.

Since philosophers can only grab their two forks simultaneously, the
system is deadlock-free for any number $n\geq2$ of philosophers. An
automatic proof requires to compute an invariant, and prove that it
has an empty intersection with the set of deadlock configurations
defined by the \wss{\kappa} formula  
\begin{align}
\mathit{deadlock}(X_w,X_e,X_f,X_b) =  \forall i ~.~  
&  [\neg X_w(i) \vee \neg X_f(i) \vee \neg X_f(\succ(i))]  \; \wedge  \nonumber \\
&  [\neg X_e(i) \vee \neg X_b(i) \vee \neg X_b(\succ(i))]
\label{deadlock-free}
\end{align}
  \noindent where $X_w$, $X_e$, $X_f$, $X_b$ are set variables, the intended meaning 
  of $X_w(i)$ resp. $X_e(i)$ is that the $i$-th philosopher is waiting, resp. eating, and 
  the intended meaning of $X_f(i)$ resp. $X_b(i)$ is that the $i$-th fork is free, resp. busy.  
Our method automatically computes from $\interform_{\mathit{philo}}$ the formula  
\begin{align}
\mathit{\trappred}(X_w,X_e,X_f,X_b) = \forall i ~.~  
& [X_w(i) \vee X_f(i) \vee X_f(\succ(i))] \;\;  \leftrightarrow  \nonumber \\
& [X_e(i) \vee X_b(i) \vee X_b(\succ(i))] \enspace.
\label{trapinv}
\end{align}
Contrary to other approaches, the computation does not require any
state-space exploration. The method guarantees that its set of models
is an inductive invariant. Since the conjunction 
$\mathit{deadlock} \wedge \trappred$ is
unsatisfiable, which can be automatically checked, the system is
deadlock-free, for any number of philosophers.

\section{Parameterized Component-based Systems}

A \emph{component type} is a tuple $\acomptype =
\tuple{\ports, \states, \initstate, \rules}$, where $\ports =
\set{p,q,r,\ldots}$ is a finite set of \emph{ports}, $\states$ is a
finite set of \emph{states}, $\initstate\in\states$ is an initial
state and $\rules \subseteq \states \times \ports \times \states$ is a
set of \emph{transitions} denoted $s \arrow{p}{} s'$, for $s,s' \in
\states$ and $p \in \ports$. We assume there are no two different transitions with the
same port. 

A \emph{component-based system} $\asys =
\tuple{\typeno{\acomptype}{1}, \ldots,
  \typeno{\acomptype}{N},\interform}$ consists of a fixed number
$N\geq1$ of component types $\typeno{\acomptype}{k} =
\tuple{\typeno{\ports}{k}, \typeno{\states}{k},
  \typeno{\initstate}{k}, \typeno{\rules}{k}}$ and an
\emph{interaction formula} $\interform$. In the dining philosophers
there are two component types, $\mathsf{Philosopher}$ and
$\mathsf{Fork}$, each with two states and two transitions, as shown in
Fig. \ref{fig:philosophers}.  We assume that $\typeno{\ports}{i} \cap
\typeno{\ports}{j} = \emptyset$ and $\typeno{\states}{i} \cap
\typeno{\states}{j} = \emptyset$, for all $1 \leq i < j \leq N$. We
denote the component type of a port $p$ or a state $s$ by $\typeof{p}$
and $\typeof{s}$, respectively.  For instance, in
Fig. \ref{fig:philosophers} we have $\typeof{p} = \typeof{g} =
\typeof{w} = \typeof{e} = \mathsf{Philosopher}$ and $\typeof{t} =
\typeof{\ell} = \typeof{f} = \typeof{b} = \mathsf{Fork}$.

The interaction formula $\interform$ determines the family of systems
we can construct out of these components.  It does so by specifying,
for each possible number of replicated instances (for example, 3
philosophers and 3 forks), which are the possible interactions between
them. An interaction consists of a set of transitions that are
executed simultaneously. For example, in an interaction philosopher 3
executes a $g$(et) transition simultaneously with $t$(ake) transitions
of the forks 2 and 3. Before formalizing this, we introduce the syntax
and semantics of Interaction Logic.

\smallskip
\paragraph{Interaction Logic.} For a constant $\kappa\geq1$, fixed throughout the paper,
the \emph{Interaction Logic} \il{\kappa} is built on top of a
countably infinite set $\vars$ of variables, the set $\preds =
\bigcup_{k=1}^N \typeno{\ports}{k}$ of monadic predicate symbols
ranged over by $\apred$ (i.e.\ the logic has a predicate symbol for
each port), the binary predicate $\leq$, and the \emph{successor}
functions $\succ_0, \ldots, \succ_{\kappa-1}$, of arity one.  The
formulae of \il{\kappa} are generated by the syntax
\[\begin{array}{rclr}
t & := & i \in \vars \mid \succ_0(t) \mid \ldots \mid \succ_{\kappa-1}(t) & \text{ terms} \\
\phi & := & t_1 \leq t_2 \mid \apred(t) \mid \phi_1 \wedge \phi_2 \mid \neg\phi_1 \mid \exists i ~.~ \phi_1 & \text{ formulae}
\end{array}\]
Abbreviations like $t_1 = t_2$, $t_1 < t_2$, $\phi_1 \vee \phi_2$,
$\phi_1 \leftrightarrow \phi_2$, and $\forall i ~.~ \phi$ are defined
as usual. \il{\kappa}\ is interpreted over finite ranked trees of
arity $\kappa$, which we identify with a prefix-closed language of
words, also called \emph{nodes}, over the alphabet $\{0, \ldots,
\kappa -1 \}$. The root of the tree is the empty word $\epsilon$, and
the children of $w$ are $w0, w1, \ldots, w{(\kappa-1)}$.  Formally, an
\emph{interpretation} or \emph{structure} is a pair $\I=(\Universe,\iota)$,
where the universe $\Universe$ is a tree and $\iota$ assigns a node to each
variable and a set of nodes to each predicate in $\preds$. The
predicate $\leq$ and the functions $\succ_0, \ldots, \succ_{k-1}$ have
the usual fixed interpretations: If $t$ and $t'$ are interpreted as
$w$ and $w'$, then $t_1 \leq t_2$ holds if{}f $w$ is a prefix of $w'$,
and $\succ_i(t)$ is interpreted as the node $wi$, if $wi \in \Universe$, and
as the root $\epsilon$ otherwise. So, loosely speaking, successor
functions wrap around to the root.

When $\kappa =1$, formulae are interpreted on languages $\{\epsilon,
0, 00, \ldots, 0^{n-1}\}$ for some number $n$. To simplify notation,
in this case we assume that they are interpreted over the set $\{0,1,
\ldots, n-1\}$, and $\succ_0$ is the usual successor function on
numbers, modulo $n$. 

Intuitively, a universe $\Universe$ determines an instance of the
component-based system, with one instance of each component for each
$w \in \Universe$.  So, for example, for $\kappa = 1$ and $\Universe=\{0,1,2,
\ldots, n-1\}$ we have in our running example philosophers $0, 1,
\ldots, n-1$ and forks $0, 1, \ldots, n-1$. Generally, with $\kappa=1$
we can describe pipeline and token-ring architectures, whereas higher
values describe tree-shaped architectures.

\smallskip
\paragraph{Interaction formulae.}
A formula of \il{\kappa} is an \emph{interaction formula} if it is
the conjunction of the following formula:
\begin{equation}\label{eq:inter-axiom}
  \begin{array}{c}
  \forall i \forall j ~.~ \bigwedge_{\begin{array}{c}
      \scriptstyle{p,q \in \preds} \\[-2mm]
      \scriptstyle{\typeof{p} = \typeof{q}}
  \end{array}} p(i) \wedge q(j) \rightarrow i \neq j
  \end{array}
\end{equation}
with a finite disjunction of formulae of the form: 
\begin{equation}\label{eq:param-interform}
 \begin{array}{c}
   \mathfrak{C}(i_1, \ldots,  i_\ell)  \isdef  \;\;  \varphi \;\; 
   \wedge \;\; \bigwedge_{j=1}^\ell p_j(i_j) \;\; \wedge \;\;
   \bigwedge_{j=1}^{m} \forall k ~.~ \psi_j \rightarrow
   q_j(k)
 \end{array}
\end{equation}
\noindent where $\varphi,\psi_{1}, \ldots, \psi_{m}$ are
conjunctions of atomic formulae of the form $t_1 \leq t_2$ and their
negations. Intuitively, formula (\ref{eq:inter-axiom}) is a generic axiom that prevents
two ports of the same instance of a component type from
interacting. The formulae of form (\ref{eq:param-interform}) are called 
 the \emph{clauses} of the interaction formula.

\begin{example}
Consider a
component-based system $\asys = \tuple{\typeno{\acomptype}{1}, \typeno{\acomptype}{2}, \interform}$,
where $\typeno{\acomptype}{1}$
and $\typeno{\acomptype}{2}$ have ports $p_1$ and $p_2$, respectively, 
and $\interform$ has one single clause
$$\mathfrak{C}(i, j, k) = \;\; (i < j  \wedge k = \succ(j)) \;\;
\wedge \;\; (p_1(i) \wedge p_2(j)) \;\; \wedge \;\; \forall i . i > k \rightarrow p_1(i)$$
\noindent  $\interform$ states that an interaction consists of: the $i$-th process of type
$\typeno{\acomptype}{1}$ executes transition $p_1$; the $j$-th process
of type $\typeno{\acomptype}{2}$ executes $p_2$; and, for every $i >
(j+1) \mod n$, the $i$-th process of type $\typeno{\acomptype}{1}$
executes transition $p_1$ as well; all this happens simultaneously in
one atomic step. \hfill$\blacksquare$
\end{example} 

Loosely speaking, (\ref{eq:param-interform}) states that in an
interaction $\ell$ components can simultaneously engage in a
multiparty rendez-vous, together with a broadcast to the ports $q_{1},
\ldots, q_{m}$ of the components whose indices satisfy the constraints
$\psi_{1}, \ldots, \psi_{m}$, respectively. An example of peer-to-peer
rendez-vous with no broadcast is the dining philosophers system in
Fig. \ref{fig:philosophers}, whereas examples of broadcast are found
among the test cases in \S\ref{sec:experiments}. In the next section
we show that, despite this generality, it is possible to construct a
trap invariant for any interaction formula in a purely syntactic way.



Observe that the formula does not explicitly specify that every other
process remains idle. Formally, as we will see in the next section,
the system has an interaction for each \emph{minimal model} of
(\ref{eq:param-interform}), which allows us not to have to specify
idleness.  Given structures $\I_1 = (\Universe,\iota_1)$ and $\I_2 =
(\Universe,\iota_2)$ sharing the same universe $\Universe$, we say $\I_1 \strord
\I_2$ if and only if $\iota_1(\apred) \subseteq \iota_2(\apred)$ for
every $\apred \in \preds$.  Given a formula $\phi$, a structure $\I$
is a \emph{minimal model} of $\phi$ if $\I \models \phi$ and, for all
structures $\I'$ such that $\I' \strord \I$ and $\I'\neq\I$, we have
$\I' \not\models \phi$.

\subsection{Execution Semantics of Component-based Systems}

The semantics of a component-based system $\asys = \tuple{\typeno{\acomptype}{1}, \ldots,
  \typeno{\acomptype}{N},\interform}$ is an infinite
family of Petri Nets, one for each universe of $\Gamma$.
The reachable markings and actions of the Petri Net characterize the reachable global
states and transitions of the system, respectively. To fix notations,
we recall several basic definitions. 

\smallskip
\paragraph{Preliminaries: Petri Nets.} A \emph{Petri Net} (PN) is a tuple $\anet =
\tuple{\places,\trans,\edges}$, where $\places$ is a set of
\emph{places}, $\trans$ is a set of \emph{transitions}, $\places \cap
\trans = \emptyset$, and $\edges \subseteq (\places \times \trans) \cup
(\trans \times \places)$ is a set of \emph{edges}. The elements of
$\places \cup \trans$ are called \emph{nodes}. Given nodes $x,y \in
\places \cup \trans$, we write $E(x,y)\isdef1$ if $(x,y) \in E$ and
$E(x,y)\isdef0$, otherwise. For a node $x$, let $\pre{x} \isdef \set{y
  \in \places \cup \trans \mid E(y,x)=1}$, $\post{x} \isdef \set{y \in
  \places \cup \trans \mid E(x,y)=1}$ and lift these definitions to
sets of nodes.

A \emph{marking} of $\anet$ is a function $\amark : \places
\rightarrow \nat$. A transition $t$ is \emph{enabled} in $\amark$ if
and only if $\amark(s) > 0$ for each place $s \in \pre{t}$. 
For all markings $\amark$, $\amark'$ and transitions $t$, we write $\amark
\arrow{t}{} \amark'$ whenever $t$ is enabled in $\amark$ and
$\amark'(s) = \amark(s) - E(s,t) + E(t,s)$, for all $s \in
\places$. Given two markings $\amark$ and $\amark'$, a finite sequence
of transitions $\sigma = t_1, \ldots,t_n$ is a \emph{firing sequence},
written $\amark \arrow{\sigma}{} \amark'$ if and only if
either \begin{inparaenum}[(i)]
\item $n=0$ and $\amark=\amark'$, or
\item $n\geq1$ and there exist markings $\amark_1, \ldots,
  \amark_{n-1}$ such that $\amark \arrow{t_1}{} \amark_1 \ldots
  \amark_{n-1} \arrow{t_{n}}{} \amark'$.
\end{inparaenum}

A \emph{marked Petri Net} is a pair $\amarkednet=(\anet,\amark_0)$,
where $\amark_0$ is the \emph{initial marking} of $\anet$. A marking
$\amark$ is \emph{reachable} in $\amarkednet$ if there exists a firing
sequence $\sigma$ such that $\amark_0 \arrow{\sigma}{} \amark$. We
denote by $\reach{\amarkednet}$ the set of reachable markings of
$\amarkednet$. A marked PN $\amarkednet$ is \emph{$1$-safe} if
$\amark(s) \leq 1$, for each $s \in \places$ and $\amark \in
\reach{\amarkednet}$. All PNs considered in the following will be
1-safe and we shall silently blur the distinction between a marking
$\amark : \places \rightarrow \set{0,1}$ and the boolean valuation
$\beta_\amark : \places \rightarrow \set{\bot,\top}$ defined as
$\beta_\amark(s)=\top \iff \amark(s)=1$.  A set of markings
$\mathcal{M}$ is an \emph{inductive invariant} of
$\amarkednet=(\anet,\amark_0)$ if and only if $\amark_0 \in
\mathcal{M}$ and for each $\amark \arrow{t}{} \amark'$ such that
$\amark \in \mathcal{M}$, we have $\amark' \in \mathcal{M}$.

\smallskip
\paragraph{Petri Net Semantics of Component-Based Systems.}
We define the semantics of a component-based system as an infinite
family of 1-safe Petri Nets.  Let $\asys =
\tuple{\typeno{\acomptype}{1}, \ldots,
  \typeno{\acomptype}{N},\interform}$ be a system with component types
$\typeno{\acomptype}{k} = \tuple{\typeno{\ports}{k},
  \typeno{\states}{k}, \typeno{\initstate}{k}, \typeno{\rules}{k}}$,
for every $k=1,\ldots,N$.  Fix a universe $\Universe$ of $\interform$. We
define a marked Petri Net $\amarkednet^{\Universe}_\asys \isdef
(\tuple{\places,\trans,\edges}, \amark_0)$ as
follows: \begin{compactitem}
\item $\places \isdef \left( \bigcup_{k=1}^N \typeno{\states}{k}
  \right) \times \Universe$ that is, the net has a place $(s, u)$ for each
  state $s$ of each component type, and for each node $u$.
\item For each minimal model $\I = (\Universe,\iota)$ of a clause
  $\mathfrak{C}$ of $\interform$, the set $\trans$ contains a
  transition $\atrans_\iota \in \trans$, and the set $\edges$ contains
  edges $((s,u),\atrans_\iota)$ and $(\atrans_\iota, (s',u))$ for
  every $s \arrow{p}{} s' \in \left( \bigcup_{k=1}^N
  \typeno{\rules}{k} \right)$ such that $u \in \iota(p)$. Nothing else
  is in $\trans$ or $\edges$. Intuitively, $\atrans_\iota$
  ``synchronizes'' all the transitions $s \arrow{p}{} s'$ of the
  different components occurring in the interaction.
\item For each $1 \leq k \leq N$, each $s \in \typeno{\states}{k}$ and
  each $u \in \Universe$, $\amark_0((s,u)) = 1$ if $s =
  \typeno{\initstate}{k}$ and $\amark_0((s,u)) = 0$, otherwise that
  is, $\amark_0$ contains the places $(s, u)$ such that $s$ is an
  initial state.
\end{compactitem}

It follows immediately from this definition that
$\amarkednet^{\Universe}_\asys$ is a 1-safe Petri Net.  Indeed, for every $u
\in \Universe$, for every component-type $\typeno{\acomptype}{k}$, and for
every reachable marking $m$, we have $\sum_{s \in \typeno{\states}{k}}
m((s, u)) = 1$. This reflects that the instance of
$\typeno{\acomptype}{k}$ at $u$ is always in exactly one of the states
of $\typeno{\states}{k}$; if $s$ is that state, then $(s, u)$ is the
place carrying the token.

\begin{example}
Consider our running example, with $\Universe = \{0,1, \ldots, n-1\}$, i.e.,
$n$ philosophers and $n$ forks.  Since the interaction formula
(\ref{interform}) has no constants, its models are pairs $(\Universe,
\iota)$, where $\iota$ gives the interpretation of the free variable
$i$ and the predicates $g$, $t$, etc. The first disjunct of
(\ref{interform}) is $[g(i) \wedge t(i) \wedge t(\succ(i))]$. It has a
minimal model for each $k \in \Universe$, namely the model with $\iota(i)=k$,
$\iota(g)= \{k\}$ and $\iota(t) = \{k, (k+1) \mod n\}$. In the
interaction produced by this model, the $k$-th philosopher executes
transition $g$(et), the forks with numbers $k$ and $(k+1) \mod n$
execute transition $t$(ake), and all other philosophers and forks
remain idle. The second disjunct yields the interactions in which a
philosopher puts down its forks.
\begin{figure}[ht]
\centerline{\input{philosophers-fig.tex}}
\caption{Petri Net of the dining philosophers for the universe $\Universe = \{0,1,2\}$. In reality, the 
two pink and green places are only one place.}
\label{fig:philonet}
\end{figure}
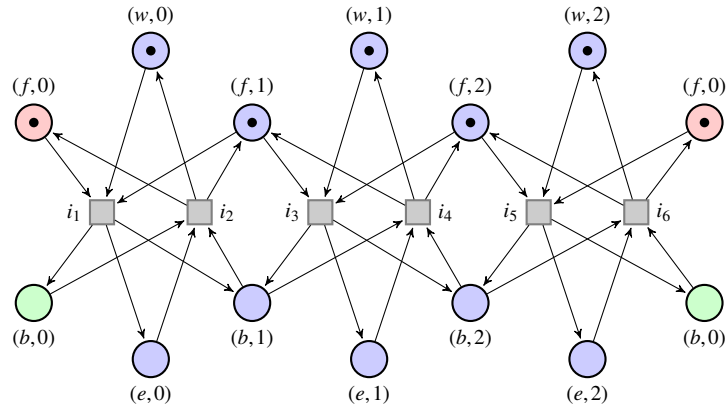
Fig. \ref{fig:philonet} shows the Petri Net $\amarkednet^{\Universe}_\asys$
for universe $\Universe = \{0,1,2\}$. For clarity, the places $(f,0)$ and $(b,0)$ have been duplicated; 
in reality the two copies are merged.
The places of each philosopher are $\{ (w,i),(e, i) \}$ for $i=0,1,2$. 
For example, transition $i_3$ corresponds to the minimal model $\{ (g,1), (t,1), (t,2) \}$, in which 
philosopher 1 takes forks 1 and 2.
\end{example}

\section{Trap Invariants}
\label{sec:trap-invariants}

Given a Petri Net $\anet = (\places, \trans,
\edges)$, a set of places $W \subseteq \places$ is called a
\emph{trap} if and only if $\post{W} \subseteq \pre{W}$. A trap $W$ of
$\anet$ is an \emph{initially marked trap} (IMT) of the marked PN
$\amarkednet = (\anet,\amark_0)$ if and only if $\amark_0(s)=\true$
for some $s \in W$. 

\begin{example}\label{ex:traps}
The Petri Net of Fig. \ref{fig:philonet} has many traps. Examples are $\{ (f,1), (b,1) \}$
and $\{ (f,0), (b,1), (f,2), (e,2) \}$.
\end{example}

An IMT defines an invariant of the Petri Net, because some place in
the trap will always be marked, no matter which sequence of transitions is
fired. The \emph{trap invariant} of $\amarkednet$ is the set of
markings that mark each IMT of $\amarkednet$. Clearly, since marked
traps remain marked, the set of reachable markings is contained in the
trap invariant. Hence, to prove that a certain set of markings is
unreachable, it is sufficient to prove that the this set has empty
intersection with the trap invariant. For self-completeness, we
briefly discuss the computation of the trap invariant for a given
marked Petri Net of fixed size, before explaining how this can be done
for the infinite family of marked Petri Nets defining the executions
of parameterized systems.

The \emph{trap constraint} of a Petri Net $\anet = (\places,\trans,\edges)$
is the formula:
\[\begin{array}{c}
\trapconstraint{\anet} \isdef \bigwedge_{t \in T} \big(\bigvee_{\text{$x \in \pre{t}$}} x\big)
\rightarrow \big(\bigvee_{\text{$y \in \post{t}$}} y\big)
\end{array}\]
where each place $x,y \in \places$ is viewed as a propositional variable. 
It is not hard to show\footnote{See e.g.\ \cite{BarkaouiLemaire89} for
  a proof.} that any boolean valuation $\beta : \places \rightarrow
\set{\bot,\top}$ that satisfies the trap constraint
$\trapconstraint{\anet}$ defines a trap $W_\beta$ of $\anet$ in the
obvious sense $W_\beta = \set{s \in \places \mid \beta(s) =
  \top}$. Further, if $\amark_0 : \places \rightarrow \set{0,1}$ is
the initial marking of a 1-safe PN $\anet$ and $\mu_0 \isdef
\bigvee_{\amark_0(s)=1} s$ is a propositional formula, then every
valuation of $\mu_0 \wedge \trapconstraint{\anet}$
defines an IMT of $(\anet,\amark_0)$.
Usually, computing invariants requires building a sequence of
underapproximants whose limit is the least fixed point of an
abstraction of the transition relation of the system
\cite{CousotCousot79}. This is not the case of the trap
invariant, that can be directly computed from the trap constraint
and the initial marking \cite{BensalemBNS09,BozgaIosifSifakis19a}.

In the rest of the section we construct a parameterized trap constraint
that characterizes the traps, not of one single net, as
$\trapconstraint{\anet}$, but of the infinite family of Petri Nets
obtained from a component-based system. The parameterized trap
constraint is a formula of \wss{\kappa}. In Section \ref{sec:il-wss}
we first explain how to embed our interaction logic into \wss{\kappa},
and in Section \ref{sec:partrap} we construct the parameterized trap
constraint.

\subsection{From \il{\kappa}\ to \wss{\kappa}{}}
\label{sec:il-wss}

We briefly recall the syntax and semantics of \wss{\kappa}\ , the
monadic second order logic \wss{\kappa}\ of $\kappa$ successors (see
e.g. \cite{KhoussainovNerode}).  Let $\Vars$ be a countably infinite
set of second-order variables (also called set variables), denoted as
$X,Y,\ldots$ in the following. The syntax of \wss{\kappa}\ is:
\[\begin{array}{rclr}
t & := & \wssroot \mid x \mid \succ_0(t) \mid \ldots \mid \succ_\kappa(t) & \text{ terms} \\
\phi & := & t_1 = t_2 \mid \apred(t) \mid X(t) \mid
\phi_1 \wedge \phi_2 \mid \neg\phi_1 \mid \exists x ~.~ \phi_1 \mid
\exists X ~.~ \phi_1 & \text{ formulae} 
\end{array}\]
So \wss{\kappa}\ extends \il{\kappa}, with the constant symbol $\wssroot$, atoms $X(t)$ and monadic second order quantifiers $\exists X ~.~ \phi$. We can consider
w.l.o.g. equality atoms $t_1=t_2$ instead of the inequalities $t_1\leq
t_2$ in \il{\kappa}, because the latter can be defined in \wss{\kappa}
as usual:
\[\begin{array}{c}
x \leq y \isdef \forall X ~.~ \mathit{closed}(X) \wedge X(x) \rightarrow X(y) 
\hspace*{5mm} \mathit{closed}(X) \isdef \forall x ~.~ X(x) \rightarrow \bigwedge_{i=0}^{\kappa-1}X(\succ_i(x)) 
\end{array}\]

Like \il{\kappa}, the formulae of \wss{\kappa}\ are interpreted on
ordered trees of arity $\kappa$. The models of \wss{\kappa} are
structures $(\Universe, \iota)$, where $\iota$ assigns the root of the tree
to $\wssroot$, a node $\iota(x)$ to each variable $x\in\vars$ and a
set $\iota(X) \subseteq \Universe$ to each set variable $X\in\Vars$. The
satisfaction relation $(\Universe,\iota) \wsmodels \phi$ is defined as for
\il{\kappa}, with one difference: in \il{\kappa}, the successor of a
leaf of a tree is the root of the tree, while in \wss{\kappa}\ the
successor of leaf is, by convention, the leaf itself \cite[Example
  2.10.3]{KhoussainovNerode}.  This is the only reason why \il{\kappa}
is not just a fragment of \wss{\kappa}.


We define an embedding of \il{\kappa}\ formulae, without occurrences
of predicates and set variables, into \wss{\kappa}. W.l.o.g. we
consider \il{\kappa}\ formulae that have been previously flattened,
i.e\ the successor function occurs only within atomic propositions of
the form $\succ_i(x) = y$. This is done by replacing each atomic
proposition of the form $\succ_{i_1}( \ldots \succ_{i_n}(x) \ldots) =
y$ by the formula $\exists x_1 \ldots \exists
x_{n} ~.~ x_n = \succ_{i_n}(x) \wedge y = \succ_{i_1}(x_1) \wedge
\bigwedge_{j=1}^{n-1} x_j = \succ_{i_j}(x_{j+1})$.  The translation of
an \il{\kappa}\ formula $\phi$ into \wss{\kappa}\ is the formula
$\mathit{Tr}(\phi)$, defined recursively on the structure of $\phi$,
is homomorphic w.r.t. the first-order connectives and:
\[\mathit{Tr}(\succ_i(x)=y) \isdef (\neg\max(x) \wedge \succ_i(x)=y)
\vee (\max(x) \wedge y = \wssroot)\] We show that a formula $\phi$ of
\il{\kappa} and its \wss{\kappa} counterpart $\mathit{Tr}(\phi)$ are
equivalent:

\begin{restatable}{lemma}{LemmaIlWss}\label{lemma:il-wss}
  Given an \il{\kappa}\ formula $\phi$, for any structure $\I =
  (\Universe,\iota)$, we have $\I \ilmodels \phi \iff \I \wsmodels
  \mathit{Tr}(\phi)$.
\end{restatable}

\subsection{Defining Parameterized Trap Invariants in \wss{\rank}}
\label{sec:partrap}

Fix a component-based system $\asys = \tuple{\typeno{\acomptype}{1},
  \ldots, \typeno{\acomptype}{N}, \interform}$ and recall that every
universe $\Universe$ induces a Petri Net $\anet^{\Universe}_{\asys}$ whose set of
places is $\bigcup_{k=1}^N \typeno{\states}{k} \times \Universe$. For every
state $s \in \bigcup_{i=1}^N \typeno{\states}{i}$, let $\statevar{s}$
be a monadic second-order variable, and let $\overline{\statevar{}}$
be the tuple of these variables in an arbitrary but fixed order. We
define a formula $\trappred(\overline{\statevar{}})$, with
$\overline{\statevar{}}$ as set of free variables, that characterizes
the traps of the infinitely many Petri Nets $\anet^{\Universe}_{\asys}$
corresponding to $\asys$. Formally,
$\trappred(\overline{\statevar{}})$ has the following property:
\begin{quote} 
For every universe $\Universe$ and for every set $P \subseteq \bigcup_{k=1}^N
\typeno{\states}{k} \times \Universe$ of places of $\anet^{\Universe}_{\asys}$: \\
$P$ is a trap of $\anet^{\Universe}_{\asys}$ if{}f the assignment
$\statevar{q}~\mapsto ~ \{ u \in \Universe \mid (q, u) \in P \}$ satisfies
$\trappred(\overline{\statevar{}})$.
\end{quote}
\noindent Observe that every assignment to $\overline{\statevar{}}$
encodes a set of places, and vice versa. So, abusing language, we can
speak of the set of places $\overline{\statevar{}}$.

We define auxiliary predicates that capture the intersection of the
set of places $\overline{\statevar{}}$ with the pre and postset of a
transition in $\anet^{\Universe}_{\asys}$. For every clause $\aclause$ of
$\interform$, of the form
(\ref{eq:param-interform}), we define the \wss{\kappa}\ formulae:
\begin{equation*}
  \small{
    \begin{array}{l}
      \intersectspre(\overline{\statevar{}}, x_{1}, \dots, x_{\ell}) =
      \bigvee_{j=1}^\ell \statevar{\pre{p_{j}}~}(x_j) \vee
      \bigvee_{j=\ell+1}^{\ell+m} \exists x_j ~.~ \mathit{Tr}(\psi_j) \wedge
      \statevar{\pre{p_j}~}(x_j)\text{ and}\\
      \intersectspost(\overline{\statevar{}}, x_{1}, \dots, x_{\ell}) =
      \bigvee_{j=1}^\ell \statevar{\post{p_{j}}~}(x_j) \vee
      \bigvee_{j=\ell+1}^{\ell+m} \exists x_j ~.~ \mathit{Tr}(\psi_j) \wedge
      \statevar{\post{p_j}~}(x_j).
    \end{array}
  }
\end{equation*}
\noindent Now we can define $\trappred(\overline{\statevar{}})$ as the conjunction of the following
formulae, one for each clause $\aclause$ (\ref{eq:param-interform}) in $\interform$:
\begin{equation}\label{eq:predicate-trap-constraint}
  \begin{array}{r}
    \small{
      \forall x_1 \ldots \forall x_\ell ~.~ \left[
        \mathit{Tr}(\varphi) \wedge
        \intersectspre(\overline{\statevar{}}, x_1, \ldots, x_\ell)
      \right]
    }\\
    \small{
    \rightarrow
      \intersectspost(\overline{\statevar{}}, x_1, \ldots, x_\ell)
    }.
  \end{array}
\end{equation}
\noindent So, intuitively, $\trappred(\overline{\statevar{}})$ states
that for every transition of the Petri Net, if the set
$\overline{\statevar{}}$ of places intersects the preset of the
transition, then it also intersects its postset. This is the condition
for the set of places to be a trap.  Formally, we obtain: 

\begin{restatable}{lemma}{LemmaTrappred}\label{lemma:trappred}
  Given a component-based system $\asys =
  \tuple{\typeno{\acomptype}{1}, \ldots,
    \typeno{\acomptype}{N},\interform}$ and a structure $\I =
  (\Universe,\iota)$, where $\iota$ is an interpretation of the set variables
  $\overline{\statevar{}}$, the set $P = \set{\tuple{s, u} \in
    \bigcup_{k=1}^N \typeno{\states}{k} \times \Universe \mid u \in
    \iota(\statevar{s})}$ is a trap of $\anet^\Universe_\asys$ if and only if
  $(\Universe,\iota) \wsmodels \trappred(\overline{\statevar{}})$.
\end{restatable}

\paragraph{Parameterized Trap Invariants in \wss{\rank}.} 
Loosely speaking, the intended meaning of
$\trappred(\overline{\statevar{}})$ is ``the set of places
$\overline{\statevar{}}$ is a trap''. Our goal is to construct a
formula stating: ``the marking $m$ marks all initially marked traps''.

Recall that the Petri Nets obtained from component-based systems are
always 1-safe, and so a marking is also a set of places. Recall,
however, that all reachable markings have the property that they place
exactly one token in the set of places modeling the set of states of a
component (loosely speaking, the set of places of the $k$-th
philosopher is $(w, k)$ and $(e, k)$, and there is always one token in
the one or the other). So we define a formula
$\marking(\overline{\statevar{}})$ with intended meaning ``the set of
places $\overline{\statevar{}}$ is a legal marking'', and another one,
$\trapconstraintpred(\overline{\statevar{}})$ with intended meaning
``the set of places $\overline{\statevar{}}$ marks every initially
marked trap (IMT)''.

In addition to the tuple of set variables $\overline{\statevar{}}$
defined above, we consider now the ``copy'' tuple
$\overline{\statevar{}'} \isdef
\tuple{\statevar{s}'}_{s\in\typeno{\states}{i},1\leq i\leq
  N}$. Intuitively, $\overline{\statevar{}}$ and
$\overline{\statevar{}'}$ represent one set of places each.

We define a ($1$-safe) marking as a set of places that marks exactly one state of each copy of each
component:
\begin{equation*}
  \marking(\overline{\statevar{}}) =
  \forall x~.~\bigwedge\limits_{1\leq i\leq N}
  \bigvee\limits_{s\in \typeno{S}{i}}\left(
    \statevar{s}(x)\wedge\bigwedge\limits_{s'\in \typeno{S}{i}~\setminus~\set{s}}
  \neg\statevar{s'}(x)
  \right).
\end{equation*}
Secondly, we give a formula describing the intersection of two sets of places:
\begin{equation*}
  \intersection(\overline{\statevar{}}, \overline{\statevar{}'}) = \exists x ~.~
  \bigvee\limits_{s\in\bigcup_{1\leq i\leq N}\typeno{\states}{i}}(
  \statevar{s}(x)\wedge\statevar{s}'(x)).
\end{equation*}
Finally, to actually capture IMTs we need to determine if a trap is initially
marked. However, this can be easily described by the formula: 
\begin{equation*}
  \initially(\overline{\statevar{}}) = \exists x ~.~
  \bigvee\limits_{1\leq i\leq N}\statevar{\typeno{\initstate}{i}}(x).
\end{equation*}
\noindent and so we can define the \emph{trap-invariant} by the \wss{\kappa} formula:
\begin{equation}\label{eq:trap-contraint-pred}
  \begin{array}{r}
  \trapconstraintpred(\overline{\statevar{}}) =
  \forall \overline{\statevar{}'}~.~\left[\trappred(\overline{\statevar{}'})
  \wedge\initially(\overline{\statevar{}'})\right]\\
    \rightarrow \intersection(\overline{\statevar{}}, \overline{\statevar{}'}).
  \end{array}
\end{equation}
Relying on Lemma \ref{lemma:trappred} we are assured that the set
represented by $\overline{\statevar{}}$ intersects all IMTs. Further,
let $\varphi(\overline{\statevar{}})$ be any formula that defines a
set of \emph{good} global states of the component-based systems (or,
equivalently, a good set of markings of their corresponding Petri
nets), with the intuition that, at any moment during execution, the
current global state of the component-based system should be good. We
can now state the following theorem, that captures the soundness of
the verification method based on trap invariants: 

\begin{restatable}{theorem}{TrapInvariant}\label{thm:trap-invariant}
 Given a component-based system $\asys$ and a \wss{\kappa}{} formula
 $\varphi(\overline{\statevar{}})$, if the formula
  \begin{equation}\label{eq:decision-formula}
    \exists \overline{\statevar{}}~.~\marking(\overline{\statevar{}})\wedge
    \trapconstraintpred(\overline{\statevar{}})\wedge
    \neg\varphi(\overline{\statevar{}})
  \end{equation}
  is unsatisfiable, then for every universe $\Universe$, the property defined
  by the formula $\varphi(\overline{\statevar{}})$ holds in every
  reachable marking of $\amarkednet^{\Universe}_{\asys}$.
\end{restatable}

In the light of the above theorem, verifying the correctness of a
component-based system with any number of active components boils down
to deciding the satisfiability of a \wss{\kappa} formula. The latter
problem is known to be decidable, albeit in general, with
non-elementary recursive complexity. A closer look at the verification
conditions of the form (\ref{eq:decision-formula}) generated by our
method suffices to see that the quantifier alternation is finite,
which implies that the time needed to decide the (un)satisfiability of
(\ref{eq:decision-formula}) is elementary recursive. Moreover, our
experiments show that these checks are very fast (less than 1 second
on an average machine) for a non-trivial set of examples.

\section{Refining Trap Invariants}
\label{sec:refine-trap-inv}

\begin{figure}[htb]
\vspace*{-\baselineskip}
\caption{Alternating Dining Philosophers}
\label{fig:philosophers2}
\vspace*{-\baselineskip}
\begin{center}
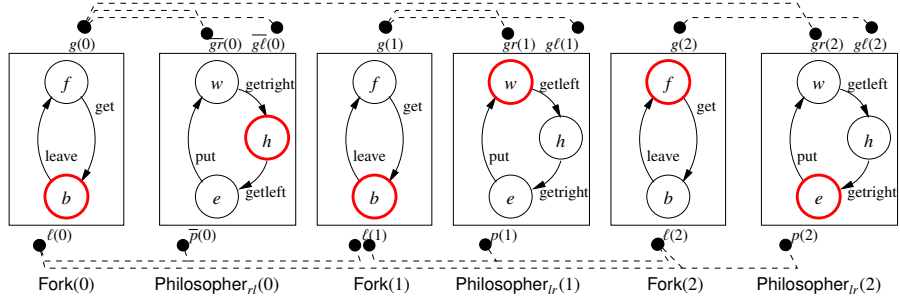
\end{center}
\vspace*{-2\baselineskip}
\end{figure}

Since the safety verification problem is undecidable for parameterized
systems \cite{AptKozen86}, the verification method based on trap
invariants cannot be complete. As an example, consider the alternating
dining philosophers system, of which an instance (for $n=3$) is shown
in Fig. \ref{fig:philosophers2}. The system consists of two
philosopher component types, namely $\mathsf{Philosopher}_{rl}$, which
takes its right fork before its left fork, and
$\mathsf{Philosopher}_{lr}$, taking the left fork before the right
one. Each philosopher has two interaction ports for taking the forks,
namely $g\ell$ (get left) and $gr$ (get right) and one port for
releasing the forks $p$ (put). The ports of the
$\mathsf{Philosopher}_{rl}$ component type are overlined, in order to
be distinguished. The $\mathsf{Fork}$ component type is the same as in
Fig. \ref{fig:philosophers}. The interaction formula for this system
$\interform_{\mathit{philo}}^{\mathit{alt}}$, shown in
Fig. \ref{fig:philosophers2}, implicitly states that only the
$0$-index philosopher component is of type
$\mathsf{Philosopher}_{rl}$, whereas all other philosophers are of
type $\mathsf{Philosopher}_{lr}$. Note that the interactions on ports
$\overline{g\ell}$, $\overline{gr}$ and $\overline{p}$ are only
allowed if $\mathit{zero}(x) \isdef \forall y ~.~ x \leq y$ holds, in
other words if $x$ is interpreted as the root of the universe
(in our case, $0$ since $\Universe=\set{0,\ldots,n-1}$).

It is well-known that any instance of the parameterized alternating
dining philosophers system consisting of at least one
$\mathsf{Philosopher}_{rl}$ and one $\mathsf{Philosopher}_{lr}$ is
deadlock-free. However, trap invariants are not enough to prove
deadlock freedom, as shown by the global state $\set{\tuple{b,0}, \tuple{h,0}, \tuple{b,1},
  \tuple{w,1}, \tuple{f,2}, \tuple{e,2}}$, marked with thick red lines in Fig.
\ref{fig:philosophers2}. Note that no interaction is enabled in this
state. Moreover, this state intersects with any trap of the marked PN
that defines the executions of this particular instance, as proved
below. Consequently, the trap invariant contains a deadlock
configuration, and the system cannot be proved deadlock-free by this
method.

\begin{restatable}{proposition}{Cex}\label{prop:cex}
  Consider an instance of the alternating dining philosophers system
  in Fig. \ref{fig:philosophers2}, consisting of components
  $\mathsf{Fork}(0)$, $\mathsf{Philosopher}_{rl}(0)$,
  $\mathsf{Fork}(1)$, $\mathsf{Philosopher}_{lr}(1)$,
  $\mathsf{Fork}(2)$ and $\mathsf{Philosopher}_{lr}(2)$ placed in a
  ring, in this order. Then each nonempty trap of this system contains
  one of the places $\tuple{b,0}, \tuple{h,0}, \tuple{b,1},
  \tuple{w,1}, \tuple{f,2}$ or $\tuple{e,2}$.
\end{restatable}

However, the configuration is unreachable by a real execution of the
PN, started in the initial configuration that marks $\tuple{f,i}$ and
$\tuple{w,i}$, for all $i=0,1,2$. An intuitive reason is that, in any reachable
configuration, each fork is in state $f$(ree) only if none of its
neighboring philosophers is in state $e$(ating). In order to prove
deadlock freedom, one must learn this and other similar
constraints. Next, we present a heuristic method for strengthening the
trap invariant, that infers such universal constraints, involving a
fixed set of components.

\subsection{One Invariants}\label{sec:flow-invariants}

As shown by the example above, trap constraints do sometimes fail to
prove interesting properties. Hence, it is desirable to refine the
overapproximation of viable markings to exclude more spurious
counterexamples. In order to do so, we consider a special class of
\emph{linear invariants}, called \emph{$1$-invariants} in the
following. Although linear invariants are not structural and rely on
the set of reachable markings of a marked Petri Net, the set of
$1$-invariants can be sufficiently under-approximated by structural
conditions. 

\begin{definition}\label{def:1-invariants}
Given a marked PN $\amarkednet =((\places, \trans,
    \edges), \amark_0)$, with $\places = \set{s_1, \ldots, s_n}$, a
vector $\vec{a} = (a_1, \ldots, a_n) \in \set{0,1}^n$ is a
\emph{$1$-invariant} of $\amarkednet$ if and only if, for each
reachable marking $\amark \in \reach{\amarkednet}$, we have
$\sum_{i=1}^n a_i \cdot \amark(s_i) = 1$. 
\end{definition}


The following lemma states the structural properties that sufficiently
define $1$-invariants. However, the opposite is not true: the are
$1$-invariants not captured by these conditions. Taking the
intersection of this set of $1$-invariants defines a weaker invariant,
for the net, which is sound for our verification purposes.  

 \begin{restatable}{lemma}{OneInvariant}\label{lemma:one-invariant}
   Given a marked PN $\amarkednet = ((\states, \trans,
       \edges), \amark_0)$, a set of places $\flow \subseteq \states$
   is a $1$-invariant if the following hold: 
   \begin{compactenum}
   \item\label{it1:one-invariant} $\sum\limits_{s \in
     \flow}\amark_{0}(s) = 1$,
   \item\label{it2:one-invariant} either $\card{\flow \cap \pre{t}~} =
     \card{\flow \cap \post{t}~} = k$ with $k \in \set{0, 1}$ or
     $\card{\flow \cap \pre{t}~} > 1$ for every $t \in \trans$.
   \end{compactenum}
 \end{restatable}

We devote the rest of this section to describe \wss{\kappa} formulae
which capture the structural properties necessary to define
$1$-invariants as laid down by Lemma \ref{lemma:one-invariant}
(\ref{it2:one-invariant}). As demonstrated in Section
\ref{sec:trap-invariants} the pre- and postset of transitions, as well
as general sets of places in a PN describing the execution semantics
can be defined in \wss{\kappa}. Hence, we present the definitions of
the following formulae only in the appendix and just give the
intuitions here. 

As before, we fix two tuples of set variables $\overline{\statevar{}}$
and $\overline{\statevar{}'}$, with one variable $X_s$ for each state
$s \in \bigcup_{i=1}^N \typeno{\states}{i}$ and we define the formulae: 
\begin{compactitem}
\item $\uniqueinitially(\overline{\statevar{}})$ which captures that the
  set of places induced by an interpretation of $\overline{\statevar{}}$
  uniquely intersects the set of all initial states, and
\item $\uniqueintersection(\overline{\statevar{}}, \overline{\statevar{}'})$
  which states that the set of places induced by an interpretation of
  $\overline{\statevar{}}$ and $\overline{\statevar{}'}$ intersect uniquely.
\end{compactitem}
Given a transition $t$ of the marked Petri Net $\amarkednet^\Universe_\asys$
defining the execution semantics of a component-based system $\asys$,
for a universe $\Universe$, we consider the following formulae: \begin{compactitem}
\item $\uniquepre(\overline{\statevar{}}, x_{1}, \ldots, x_{\ell})$ which
  describes that the set of places encoded by the interpretation of
  $\overline{\statevar{}}$ uniquely intersects $\pre{t}$ and
\item $\uniquepost(\overline{\statevar{}}, x_{1}, \ldots, x_{\ell})$
  which in the same sense captures the unique intersection with
  $\post{t}$.
\end{compactitem}
Now we define a predicate $\flowpred$ which consists of a conjunction of
$\uniqueinitially$ and the formulae: 
\begin{equation}
  \small{
    \begin{array}{r}
      \forall x_{1}, \ldots, \forall x_{\ell} ~.~ (
      \mathit{Tr}(\varphi) \rightarrow [
        \neg\intersectspre \wedge \neg\intersectspost\\
        \vee \uniquepre \wedge \uniquepost\\
        \vee \intersectspre \wedge \neg \uniquepre
        ]
      )
    \end{array}
  }
\end{equation}
one for each clause $\aclause$ in $\interform$. We show the soundness
of this definition, by the following: 

\begin{restatable}{lemma}{LemmaFlowPred}\label{lemma:flowpred}
  Given a component-based system $\asys =
  \tuple{\typeno{\acomptype}{1}, \ldots,
    \typeno{\acomptype}{N},\interform}$ and a tuple of set variables
  $\overline{\statevar{}}$, one for each state in a component of
  $\asys$. Then, for any structure $(\Universe,\iota)$, such that $\iota$
  interprets the variables in $\overline{\statevar{}}$, the set $P =
  \set{\tuple{s,u} \in \bigcup_{i=1}^N \typeno{\states}{i} \times \Universe
    \mid u \in \iota(\statevar{s})}$ is a $1$-invariant of
  $\amarkednet^\Universe_\asys$ if $(\Universe, \iota) \wsmodels
  \flowpred(\overline{\statevar{}})$.
\end{restatable}
We may now define the \emph{$1$-invariant} analogously to the
trap-invariant before:
\begin{equation}\label{eq:flow-contraint-pred}
  \flowinvariant(\overline{\statevar{}}) =
  \forall \overline{\statevar{}'}~.~\flowpred(\overline{\statevar{}'})
    \rightarrow \uniqueintersection(\overline{\statevar{}},
    \overline{\statevar{}'}).
\end{equation}
And by analogously reasoning we obtain a refinement of Theorem
\ref{thm:trap-invariant} since every reachable marking has to satisfy both
invariants.

\begin{restatable}{theorem}{FlowInvariant}\label{thm:flow-invariant}
 Given a component-based system $\asys$ and a \wss{\kappa} formula
 $\varphi(\overline{\statevar{}})$, if the formula:
  \begin{equation}\label{eq:decision-formula-flow}
    \exists \overline{\statevar{}}~.~\marking(\overline{\statevar{}})\wedge
    \flowinvariant(\overline{\statevar{}})\wedge
    \trapconstraintpred(\overline{\statevar{}})\wedge
    \neg\varphi(\overline{\statevar{}})
  \end{equation}
  is unsatisfiable, then for every universe $\Universe$, the property defined by
  the formula $\varphi(\overline{\statevar{}})$ holds in every
  reachable marking of $\amarkednet^{\Universe}_{\asys}$.
\end{restatable}

\section{Experiments}\label{sec:experiments}

We have implemented a prototype (called \texttt{ostrich}) of this verification
procedure to evaluate the viability of our approach. All the files and scripts
used can be found online\footnote{\url{https://gitlab.lrz.de/i7/ostrich}}.
The current version of the prototype can only handle token-ring and
pipeline topologies, but not trees; for these topologies the
verification reduces to checking satisfiability of a formula of
\wsones{}.  We have also considered one example with tree-topology
(see below), for which the formula was constructed
manually. Satisfiability of \wsones{} and \wss{2} formulae was checked
using version 1.4/17 of Mona \cite{mona}. We consider various examples
separated in categories:
\begin{description}
  \item[Cache Coherence:] Following~\cite{Delzanno00} we formalized and checked
    the described safety properties and deadlock-freedom of the following cache
    coherence protocols: Illinois, Berkeley, Synapse, Firefly, MESI, MOESI, and
    Dragon.
  \item[Mutual Exclusion:] We modelled and checked for deadlock-freedom and
    mutual exclusion Burns'~\cite{Jensen98}, Dijkstra's and Szymanski's
    \cite{AbdullaHH16} algorithms as well as a formulation of Dijkstra's
    algorithm on a ring structure with token passing~\cite{FribourgO97}.
    Furthermore, we check synchronization via a semaphore which is atomically
    aquired and by broadcasting to ensure everyone else is not in the critical
    section.
  \item[Dining Philosophers:] This is the classical problem of dining
    philosophers which all take first the right fork and then the left fork.
    We consider the following \enquote{flavors} of this problem:
    \begin{itemize}
      \item there is one philosopher who takes first her left and then her
        right fork,
      \item as above but the forks remember whom took them, and
      \item there are two global forks everyone grabs in the same order.
    \end{itemize}
  \item[Preemptive Tasks:] There are tasks which can be either waiting,
    ready, executing or preempted. Initially one task is executing while all
    others are waiting. At any point a task may become ready and any ready task
    may preempt the currently executing task. Upon finishing the executing task
    re-enables one preempted task.
    Here, we have additionally two alternatives: Firstly, we consider the case
    where always the agent with highest index resumes execution. Secondly, we
    let the processes establish the initial condition from a position where
    everyone is waiting.
  \item[Dijkstra-Scholten:] This is an algorithm that is used to detect
    termination of distributed systems by message passing along a
    tree~\cite{DijkstraS80}. Since the prototype only supports linear
    topologies we can generate the necessary formula automatically only for
    this case.
  \item[Herman:] This algorithm implements self-stabilizing token passing in
    rings. The formulation as well as all following follow
    \cite{ChenHongLinRummer17}.
  \item[Israeli-Jalfon:] This is another self-stabilizing token passing
    algorithm in rings.
  \item[Lehmann-Rabin:] This is a randomized solution to the dining
    philosophers problem.
  \item[Dining Cryptographers:] This time a group of cryptographers want to
    determine if one of them paid for a meal or a stranger.
\end{description}

The results are shown in Table \ref{tab:experiments}. The first column
reports which properties could ($\checkmark$) and could not be
verified ($\times$) because the conjunction of
trap and one-invariant was not strong enough to prove the given
property. For the consistency properties of cache-coherence protocols we give
($x$/$y$) which reads: \enquote{from $y$ properties $x$ could be established}.
The following column reports the time (in second) it takes
to prove all considered properties. To understand the next four
columns, recall that Mona constructs for a given formula $\phi(X)$ a
finite automaton recognizing all the sets $X$ for which $\varphi$
holds. Since the automata can have very large alphabets, the
transition relation of the automaton is encoded as a binary decision
diagram (BDD). The columns report the sizes of the automata (in the
format: number of states / number of nodes of the BDD) for different
formulas. More precisely, the columns trap, trap-inv, flow and
flow-inv give the sizes of the automata for $\trappred \wedge
\initially$, $\trapconstraintpred$, $\flowpred$ and $\flowinvariant$
respectively. We write \enquote{n.a.}  (for \enquote{not available})
to indicate that Mona timed out before the automaton was computed.

\begin{table}
  \input experiment-results
  \caption{Experimental results of a prototype implementation.}
  \label{tab:experiments}
\end{table}

The first observation is that the satisfiability checks often can be done
in very short time. This is surprising, because the
formulas to be checked, namely (\ref{eq:decision-formula}) and
(\ref{eq:decision-formula-flow}), exhibit one quantifier alternation
(recall that $\trapconstraintpred$ and $\flowinvariant$ contain
universal quantifiers). More specifically, since $\trapconstraintpred$
is obtained by universally quantifying over $\trappred \wedge
\initially$, one would expect the automaton for the former to be much
larger than the one for the latter, at least in some cases. But this
does not happen: In fact, the automaton for $\trapconstraintpred$ is
almost always smaller. Similarly, there is no blowup between $\flowpred$
and $\flowinvariant$. A possible explanation could be that the exponential
blowup caused by universal quantification in \wss{\kappa}~ manifests
only on theoretical corner cases, which do not occur in our examples.

%% file: philosophers.pdf_t
\begin{picture}(0,0)%
\includegraphics{philosophers.pdf}%
\end{picture}%
\setlength{\unitlength}{2960sp}%
\begingroup\makeatletter\ifx\SetFigFont\undefined%
\gdef\SetFigFont#1#2#3#4#5{%
  \reset@font\fontsize{#1}{#2pt}%
  \fontfamily{#3}\fontseries{#4}\fontshape{#5}%
  \selectfont}%
\fi\endgroup%
\begin{picture}(4599,3249)(889,-2548)
\put(4276, 89){\makebox(0,0)[b]{\smash{{\SetFigFont{9}{10.8}{\rmdefault}{\mddefault}{\updefault}{\color[rgb]{0,0,0}$\ell(\succ(k))$}%
}}}}
\put(3076,-436){\makebox(0,0)[b]{\smash{{\SetFigFont{9}{10.8}{\rmdefault}{\mddefault}{\updefault}{\color[rgb]{0,0,0}$w$}%
}}}}
\put(3076,-1636){\makebox(0,0)[b]{\smash{{\SetFigFont{9}{10.8}{\rmdefault}{\mddefault}{\updefault}{\color[rgb]{0,0,0}$e$}%
}}}}
\put(3451,-1036){\makebox(0,0)[b]{\smash{{\SetFigFont{9}{10.8}{\rmdefault}{\mddefault}{\updefault}{\color[rgb]{0,0,0}$g$}%
}}}}
\put(2701,-1036){\makebox(0,0)[b]{\smash{{\SetFigFont{9}{10.8}{\rmdefault}{\mddefault}{\updefault}{\color[rgb]{0,0,0}$p$}%
}}}}
\put(2776, 89){\makebox(0,0)[b]{\smash{{\SetFigFont{9}{10.8}{\rmdefault}{\mddefault}{\updefault}{\color[rgb]{0,0,0}$p(k)$}%
}}}}
\put(3376, 89){\makebox(0,0)[b]{\smash{{\SetFigFont{9}{10.8}{\rmdefault}{\mddefault}{\updefault}{\color[rgb]{0,0,0}$g(k)$}%
}}}}
\put(4726,-436){\makebox(0,0)[b]{\smash{{\SetFigFont{9}{10.8}{\rmdefault}{\mddefault}{\updefault}{\color[rgb]{0,0,0}$f$}%
}}}}
\put(4726,-1636){\makebox(0,0)[b]{\smash{{\SetFigFont{9}{10.8}{\rmdefault}{\mddefault}{\updefault}{\color[rgb]{0,0,0}$b$}%
}}}}
\put(5101,-1036){\makebox(0,0)[b]{\smash{{\SetFigFont{9}{10.8}{\rmdefault}{\mddefault}{\updefault}{\color[rgb]{0,0,0}$t$}%
}}}}
\put(4351,-1036){\makebox(0,0)[b]{\smash{{\SetFigFont{9}{10.8}{\rmdefault}{\mddefault}{\updefault}{\color[rgb]{0,0,0}$\ell$}%
}}}}
\put(3076,-2086){\makebox(0,0)[b]{\smash{{\SetFigFont{9}{10.8}{\rmdefault}{\mddefault}{\updefault}{\color[rgb]{0,0,0}$\mathsf{Philosopher}(k)$}%
}}}}
\put(1501,-2086){\makebox(0,0)[b]{\smash{{\SetFigFont{9}{10.8}{\rmdefault}{\mddefault}{\updefault}{\color[rgb]{0,0,0}$\mathsf{Fork}(k)$}%
}}}}
\put(4801,-2086){\makebox(0,0)[b]{\smash{{\SetFigFont{9}{10.8}{\rmdefault}{\mddefault}{\updefault}{\color[rgb]{0,0,0}$\mathsf{Fork}((k+1)\mod N)$}%
}}}}
\put(3301,-2461){\makebox(0,0)[b]{\smash{{\SetFigFont{11}{13.2}{\rmdefault}{\mddefault}{\updefault}{\color[rgb]{0,0,0}$\interform_{\mathit{philo}} = (g(i) \wedge t(i)\wedge t(\succ(i))) \vee (p(i) \wedge \ell(i) \wedge \ell(\succ(i)))$}%
}}}}
\put(1501,-436){\makebox(0,0)[b]{\smash{{\SetFigFont{9}{10.8}{\rmdefault}{\mddefault}{\updefault}{\color[rgb]{0,0,0}$f$}%
}}}}
\put(1501,-1636){\makebox(0,0)[b]{\smash{{\SetFigFont{9}{10.8}{\rmdefault}{\mddefault}{\updefault}{\color[rgb]{0,0,0}$b$}%
}}}}
\put(1876,-1036){\makebox(0,0)[b]{\smash{{\SetFigFont{9}{10.8}{\rmdefault}{\mddefault}{\updefault}{\color[rgb]{0,0,0}$t$}%
}}}}
\put(1126,-1036){\makebox(0,0)[b]{\smash{{\SetFigFont{9}{10.8}{\rmdefault}{\mddefault}{\updefault}{\color[rgb]{0,0,0}$\ell$}%
}}}}
\put(1201, 89){\makebox(0,0)[b]{\smash{{\SetFigFont{9}{10.8}{\rmdefault}{\mddefault}{\updefault}{\color[rgb]{0,0,0}$\ell(k)$}%
}}}}
\put(1801, 89){\makebox(0,0)[b]{\smash{{\SetFigFont{9}{10.8}{\rmdefault}{\mddefault}{\updefault}{\color[rgb]{0,0,0}$t(k)$}%
}}}}
\put(5176, 89){\makebox(0,0)[b]{\smash{{\SetFigFont{9}{10.8}{\rmdefault}{\mddefault}{\updefault}{\color[rgb]{0,0,0}$t(\succ(k))$}%
}}}}
\end{picture}%

%% file: philosophers-fig.tex
\begin{tikzpicture}[node distance=1.3cm,>=stealth',bend angle=45,every node/.style={scale=0.8}]

  \tikzstyle{place}=[circle,thick,draw=black,fill=blue!20,minimum size=6mm]
  \tikzstyle{red place}=[place,draw=black,fill=red!20]
  \tikzstyle{green place}=[place,draw=black,fill=green!20]
  \tikzstyle{transition}=[rectangle,thick,draw=black!50,
  			  fill=black!20,minimum size=4mm]

  \tikzstyle{every label}=[black]

    \def\d{0.6}
    \node [place,tokens=1] (w0) [label=above:{$(w,0)$}]                       			{};
    \node [place,tokens=1] (w1) [right = 4*\d  of w0, label=above:{$(w,1)$}]    	{};
    \node [place,tokens=1] (w2) [right = 4*\d  of w1, label=above:{$(w,2)$}]    	{};
    \node [place] (e0)  [below = 6*\d of w0, label=below:{$(e,0)$}] 				{};
    \node [place] (e1)  [below = 6*\d of w1, label=below:{$(e,1)$}] 				{};   
    \node [place] (e2)  [below = 6*\d of w2, label=below:{$(e,2)$}] 				{};   
    \node [red place,tokens=1] (f0)   [below left = \d and 2*\d of w0, label=above:{$(f,0)$}] 	{};
    \node [green place] (b0)   [below left = 5*\d and 2*\d of w0, label=below:{$(b,0)$}]			{};
    \node [place,tokens=1] (f1)   [below left = \d and 2*\d of w1, label=above:{$(f,1)$}]		{};
    \node [place] (b1)   [below left = 5*\d and 2*\d of w1, label=below:{$(b,1)$}] 				{};
    \node [place,tokens=1] (f2)   [below left = \d and 2*\d of w2, label=above:{$(f,2)$}]		{};
    \node [place] (b2)   [below left = 5*\d and 2*\d of w2, label=below:{$(b,2)$}] 				{};
    \node [red place,tokens=1] (f0d)   [below right = \d and 2*\d of w2, label=above:{$(f,0)$}]	{};
    \node [green place] (b0d)   [below right = 5*\d and 2*\d of w2, label=below:{$(b,0)$}]		{};
    
    \node [transition] (i1) [below left = 3*\d and 0.5*\d of w0, label=left:{$i_1$}] {}
      edge [pre]			(w0)
      edge [pre]			(f0)
      edge [pre]			(f1)
      edge [post]			(e0)
      edge [post]			(b0)
      edge [post]			(b1);
      
      \node [transition] (i2) [below right = 3*\d and 0.5*\d of w0, label=right:{$i_2$}] {}
      edge [post]			(w0)
      edge [post]			(f0)
      edge [post]			(f1)
      edge [pre]			(e0)
      edge [pre]			(b0)
      edge [pre]			(b1);
      
      \node [transition] (i3) [below left = 3*\d and 0.5*\d of w1, label=left:{$i_3$}] {}
      edge [pre] 			(w1)
      edge [pre]			(f2)
      edge [pre]			(f1)
      edge [post]			(e1)
      edge [post]			(b2)
      edge [post]			(b1);
      
      \node [transition] (i4) [below right = 3*\d and 0.5*\d of w1, label=right:{$i_4$}] {}
      edge [post]			(w1)
      edge [post]			(f2)
      edge [post]			(f1)
      edge [pre]   			(e1)
      edge [pre]			(b2)
      edge [pre]			(b1);
      
      \node [transition] (i5) [below left = 3*\d and 0.5*\d of w2, label=left:{$i_5$}] {}
      edge [pre]			(w2)
      edge [pre]			(f2)
      edge [pre]			(f0d)
      edge [post]			(e2)
      edge [post]			(b2)
      edge [post]			(b0d);
      
      \node [transition] (i6) [below right = 3*\d and 0.5*\d of w2, label=right:{$i_6$}] {}
      edge [post]			(w2)
      edge [post]			(f0d)
      edge [post]			(f2)
      edge [pre]   			(e2)
      edge [pre]			(b0d)
      edge [pre]			(b2);

%
\end{tikzpicture}

%% file: philosophers2.pdf_t
\begin{picture}(0,0)%
\includegraphics{philosophers2.pdf}%
\end{picture}%
\setlength{\unitlength}{2368sp}%
\begingroup\makeatletter\ifx\SetFigFont\undefined%
\gdef\SetFigFont#1#2#3#4#5{%
  \reset@font\fontsize{#1}{#2pt}%
  \fontfamily{#3}\fontseries{#4}\fontshape{#5}%
  \selectfont}%
\fi\endgroup%
\begin{picture}(9325,3933)(889,-3457)
\put(6001,-3361){\makebox(0,0)[b]{\smash{{\SetFigFont{10}{12.0}{\rmdefault}{\mddefault}{\updefault}{\color[rgb]{0,0,0}$\neg\mathit{zero}(x) \wedge [(g\ell(x) \wedge g(x)) \vee (gr(x) \wedge g(s(x))) \vee (p(x) \wedge \ell(x) \wedge \ell(s(x)))]$
}%
}}}}
\put(1501,-436){\makebox(0,0)[b]{\smash{{\SetFigFont{7}{8.4}{\rmdefault}{\mddefault}{\updefault}{\color[rgb]{0,0,0}$f$}%
}}}}
\put(1501,-1636){\makebox(0,0)[b]{\smash{{\SetFigFont{7}{8.4}{\rmdefault}{\mddefault}{\updefault}{\color[rgb]{0,0,0}$b$}%
}}}}
\put(1801,-661){\makebox(0,0)[lb]{\smash{{\SetFigFont{6}{7.2}{\rmdefault}{\mddefault}{\updefault}{\color[rgb]{0,0,0}get}%
}}}}
\put(1276,-1186){\makebox(0,0)[lb]{\smash{{\SetFigFont{6}{7.2}{\rmdefault}{\mddefault}{\updefault}{\color[rgb]{0,0,0}leave}%
}}}}
\put(3076,-436){\makebox(0,0)[b]{\smash{{\SetFigFont{7}{8.4}{\rmdefault}{\mddefault}{\updefault}{\color[rgb]{0,0,0}$w$}%
}}}}
\put(3076,-1636){\makebox(0,0)[b]{\smash{{\SetFigFont{7}{8.4}{\rmdefault}{\mddefault}{\updefault}{\color[rgb]{0,0,0}$e$}%
}}}}
\put(3601,-1036){\makebox(0,0)[b]{\smash{{\SetFigFont{7}{8.4}{\rmdefault}{\mddefault}{\updefault}{\color[rgb]{0,0,0}$h$}%
}}}}
\put(2851,-1186){\makebox(0,0)[lb]{\smash{{\SetFigFont{6}{7.2}{\rmdefault}{\mddefault}{\updefault}{\color[rgb]{0,0,0}put}%
}}}}
\put(3376,-436){\makebox(0,0)[lb]{\smash{{\SetFigFont{6}{7.2}{\rmdefault}{\mddefault}{\updefault}{\color[rgb]{0,0,0}getright}%
}}}}
\put(3376,-1561){\makebox(0,0)[lb]{\smash{{\SetFigFont{6}{7.2}{\rmdefault}{\mddefault}{\updefault}{\color[rgb]{0,0,0}getleft}%
}}}}
\put(9376,-436){\makebox(0,0)[b]{\smash{{\SetFigFont{7}{8.4}{\rmdefault}{\mddefault}{\updefault}{\color[rgb]{0,0,0}$w$}%
}}}}
\put(9376,-1636){\makebox(0,0)[b]{\smash{{\SetFigFont{7}{8.4}{\rmdefault}{\mddefault}{\updefault}{\color[rgb]{0,0,0}$e$}%
}}}}
\put(9901,-1036){\makebox(0,0)[b]{\smash{{\SetFigFont{7}{8.4}{\rmdefault}{\mddefault}{\updefault}{\color[rgb]{0,0,0}$h$}%
}}}}
\put(9151,-1186){\makebox(0,0)[lb]{\smash{{\SetFigFont{6}{7.2}{\rmdefault}{\mddefault}{\updefault}{\color[rgb]{0,0,0}put}%
}}}}
\put(9676,-436){\makebox(0,0)[lb]{\smash{{\SetFigFont{6}{7.2}{\rmdefault}{\mddefault}{\updefault}{\color[rgb]{0,0,0}getleft}%
}}}}
\put(9676,-1561){\makebox(0,0)[lb]{\smash{{\SetFigFont{6}{7.2}{\rmdefault}{\mddefault}{\updefault}{\color[rgb]{0,0,0}getright}%
}}}}
\put(1666,-11){\makebox(0,0)[b]{\smash{{\SetFigFont{6}{7.2}{\rmdefault}{\mddefault}{\updefault}{\color[rgb]{0,0,0}$g(0)$}%
}}}}
\put(3166,-11){\makebox(0,0)[b]{\smash{{\SetFigFont{6}{7.2}{\rmdefault}{\mddefault}{\updefault}{\color[rgb]{0,0,0}$\overline{gr}(0)$}%
}}}}
\put(3616,-11){\makebox(0,0)[b]{\smash{{\SetFigFont{6}{7.2}{\rmdefault}{\mddefault}{\updefault}{\color[rgb]{0,0,0}$\overline{g\ell}(0)$}%
}}}}
\put(9466,-11){\makebox(0,0)[b]{\smash{{\SetFigFont{6}{7.2}{\rmdefault}{\mddefault}{\updefault}{\color[rgb]{0,0,0}$gr(2)$}%
}}}}
\put(9916,-11){\makebox(0,0)[b]{\smash{{\SetFigFont{6}{7.2}{\rmdefault}{\mddefault}{\updefault}{\color[rgb]{0,0,0}$g\ell(2)$}%
}}}}
\put(9376,-2536){\makebox(0,0)[b]{\smash{{\SetFigFont{7}{8.4}{\rmdefault}{\mddefault}{\updefault}{\color[rgb]{0,0,0}$\mathsf{Philosopher}_{lr}(2)$}%
}}}}
\put(1501,-2536){\makebox(0,0)[b]{\smash{{\SetFigFont{7}{8.4}{\rmdefault}{\mddefault}{\updefault}{\color[rgb]{0,0,0}$\mathsf{Fork}(0)$}%
}}}}
\put(4801,-2536){\makebox(0,0)[b]{\smash{{\SetFigFont{7}{8.4}{\rmdefault}{\mddefault}{\updefault}{\color[rgb]{0,0,0}$\mathsf{Fork}(1)$}%
}}}}
\put(6226,-2536){\makebox(0,0)[b]{\smash{{\SetFigFont{7}{8.4}{\rmdefault}{\mddefault}{\updefault}{\color[rgb]{0,0,0}$\mathsf{Philosopher}_{lr}(1)$}%
}}}}
\put(3076,-2536){\makebox(0,0)[b]{\smash{{\SetFigFont{7}{8.4}{\rmdefault}{\mddefault}{\updefault}{\color[rgb]{0,0,0}$\mathsf{Philosopher}_{rl}(0)$}%
}}}}
\put(7876,-2536){\makebox(0,0)[b]{\smash{{\SetFigFont{7}{8.4}{\rmdefault}{\mddefault}{\updefault}{\color[rgb]{0,0,0}$\mathsf{Fork}(2)$}%
}}}}
\put(1426,-2011){\makebox(0,0)[b]{\smash{{\SetFigFont{6}{7.2}{\rmdefault}{\mddefault}{\updefault}{\color[rgb]{0,0,0}$\ell(0)$}%
}}}}
\put(2926,-2011){\makebox(0,0)[b]{\smash{{\SetFigFont{6}{7.2}{\rmdefault}{\mddefault}{\updefault}{\color[rgb]{0,0,0}$\overline{p}(0)$}%
}}}}
\put(9226,-2011){\makebox(0,0)[b]{\smash{{\SetFigFont{6}{7.2}{\rmdefault}{\mddefault}{\updefault}{\color[rgb]{0,0,0}$p(2)$}%
}}}}
\put(5401,-3061){\makebox(0,0)[b]{\smash{{\SetFigFont{10}{12.0}{\rmdefault}{\mddefault}{\updefault}{\color[rgb]{0,0,0}$\interform_{\mathit{philo}}^{\mathit{alt}} = \mathit{zero}(x) \wedge [(\overline{g\ell}(x) \wedge g(x)) \vee (\overline{gr}(x) \wedge g(s(x))) \vee (\overline{p}(x) \wedge \ell(x) \wedge \ell(s(x)))] ~\vee$
}%
}}}}
\put(7801,-436){\makebox(0,0)[b]{\smash{{\SetFigFont{7}{8.4}{\rmdefault}{\mddefault}{\updefault}{\color[rgb]{0,0,0}$f$}%
}}}}
\put(7801,-1636){\makebox(0,0)[b]{\smash{{\SetFigFont{7}{8.4}{\rmdefault}{\mddefault}{\updefault}{\color[rgb]{0,0,0}$b$}%
}}}}
\put(8101,-661){\makebox(0,0)[lb]{\smash{{\SetFigFont{6}{7.2}{\rmdefault}{\mddefault}{\updefault}{\color[rgb]{0,0,0}get}%
}}}}
\put(7576,-1186){\makebox(0,0)[lb]{\smash{{\SetFigFont{6}{7.2}{\rmdefault}{\mddefault}{\updefault}{\color[rgb]{0,0,0}leave}%
}}}}
\put(4726,-436){\makebox(0,0)[b]{\smash{{\SetFigFont{7}{8.4}{\rmdefault}{\mddefault}{\updefault}{\color[rgb]{0,0,0}$f$}%
}}}}
\put(4726,-1636){\makebox(0,0)[b]{\smash{{\SetFigFont{7}{8.4}{\rmdefault}{\mddefault}{\updefault}{\color[rgb]{0,0,0}$b$}%
}}}}
\put(5026,-661){\makebox(0,0)[lb]{\smash{{\SetFigFont{6}{7.2}{\rmdefault}{\mddefault}{\updefault}{\color[rgb]{0,0,0}get}%
}}}}
\put(4501,-1186){\makebox(0,0)[lb]{\smash{{\SetFigFont{6}{7.2}{\rmdefault}{\mddefault}{\updefault}{\color[rgb]{0,0,0}leave}%
}}}}
\put(6151,-436){\makebox(0,0)[b]{\smash{{\SetFigFont{7}{8.4}{\rmdefault}{\mddefault}{\updefault}{\color[rgb]{0,0,0}$w$}%
}}}}
\put(6151,-1636){\makebox(0,0)[b]{\smash{{\SetFigFont{7}{8.4}{\rmdefault}{\mddefault}{\updefault}{\color[rgb]{0,0,0}$e$}%
}}}}
\put(6676,-1036){\makebox(0,0)[b]{\smash{{\SetFigFont{7}{8.4}{\rmdefault}{\mddefault}{\updefault}{\color[rgb]{0,0,0}$h$}%
}}}}
\put(5926,-1186){\makebox(0,0)[lb]{\smash{{\SetFigFont{6}{7.2}{\rmdefault}{\mddefault}{\updefault}{\color[rgb]{0,0,0}put}%
}}}}
\put(6451,-436){\makebox(0,0)[lb]{\smash{{\SetFigFont{6}{7.2}{\rmdefault}{\mddefault}{\updefault}{\color[rgb]{0,0,0}getleft}%
}}}}
\put(6451,-1561){\makebox(0,0)[lb]{\smash{{\SetFigFont{6}{7.2}{\rmdefault}{\mddefault}{\updefault}{\color[rgb]{0,0,0}getright}%
}}}}
\put(4891,-11){\makebox(0,0)[b]{\smash{{\SetFigFont{6}{7.2}{\rmdefault}{\mddefault}{\updefault}{\color[rgb]{0,0,0}$g(1)$}%
}}}}
\put(6241,-11){\makebox(0,0)[b]{\smash{{\SetFigFont{6}{7.2}{\rmdefault}{\mddefault}{\updefault}{\color[rgb]{0,0,0}$gr(1)$}%
}}}}
\put(6691,-11){\makebox(0,0)[b]{\smash{{\SetFigFont{6}{7.2}{\rmdefault}{\mddefault}{\updefault}{\color[rgb]{0,0,0}$g\ell(1)$}%
}}}}
\put(7966,-11){\makebox(0,0)[b]{\smash{{\SetFigFont{6}{7.2}{\rmdefault}{\mddefault}{\updefault}{\color[rgb]{0,0,0}$g(2)$}%
}}}}
\put(4726,-2011){\makebox(0,0)[b]{\smash{{\SetFigFont{6}{7.2}{\rmdefault}{\mddefault}{\updefault}{\color[rgb]{0,0,0}$\ell(1)$}%
}}}}
\put(6076,-2011){\makebox(0,0)[b]{\smash{{\SetFigFont{6}{7.2}{\rmdefault}{\mddefault}{\updefault}{\color[rgb]{0,0,0}$p(1)$}%
}}}}
\put(7876,-2011){\makebox(0,0)[b]{\smash{{\SetFigFont{6}{7.2}{\rmdefault}{\mddefault}{\updefault}{\color[rgb]{0,0,0}$\ell(2)$}%
}}}}
\end{picture}%

%% file: experiment-results.tex
\newcommand\T{\rule{0pt}{2.6ex}}       
\newcommand\B{\rule[-1.2ex]{0pt}{0pt}} 

\setlength{\tabcolsep}{2.2pt}
\renewcommand{\arraystretch}{1.05}

\begin{center}
  \scalebox{0.8}{
    \begin{tabular}{p{4cm}|l|r|c|c|c|c}
      \multicolumn{1}{c|}{Benchmark} & \multicolumn{1}{c|}{Properties}
      & \begin{tabular}{c} time\\(s.) \end{tabular}
      & \begin{tabular}{c} trap\\ \#st. / \#tr. \end{tabular}
      & \begin{tabular}{c} trap-inv \\ \#st. / \#tr. \end{tabular}
      & \begin{tabular}{c} flow\\ \#st. / \#tr. \end{tabular}
      & \begin{tabular}{c} flow-inv \\ \#st. / \#tr. \end{tabular} \T\B \\
      \hline

      Berkeley &
      \begin{tabular}{ll}
        deadlock-freedom & \checkmark \\
        consistency properties & (1/3)\\
      \end{tabular}
      & 0.25 & 18 / 95 & 18 / 77 & 8 / 27 & 7 / 24 \\
      \hline

      Dragon &
      \begin{tabular}{ll}
        deadlock-freedom & \checkmark \\
        consistency properties & (6/7)\\
      \end{tabular}
      & 0.79 & 39 / 343 & 32 / 154 & 10 / 56 & 9 / 43 \\
      \hline

      Firefly &
      \begin{tabular}{ll}
        deadlock-freedom & \checkmark \\
        consistency properties & (0/4)\\
      \end{tabular}
      & 0.43 & 55 / 400 & 36 / 212 & 10 / 47 & 9 / 36 \\
      \hline

      Illinois &
      \begin{tabular}{ll}
        deadlock-freedom & \checkmark \\
        consistency properties & (2/2)\\
      \end{tabular}
      & 0.29 & 14 / 83 & 11 / 32 & 8 / 32 & 8 / 27 \\
      \hline

      MESI &
      \begin{tabular}{ll}
        deadlock-freedom & \checkmark \\
        consistency properties & (2/2)\\
      \end{tabular}
      & 0.23 & 12 / 58 & 11 / 35 & 8 / 27 & 7 / 24 \\
      \hline

      MOESI &
      \begin{tabular}{ll}
        deadlock-freedom & \checkmark \\
        consistency properties & (7/7) \\
      \end{tabular}
      & 0.41 & 20 / 143 & 16 / 58 & 8 / 31 & 7 / 28 \\
      \hline

      Synapse &
      \begin{tabular}{ll}
        deadlock-freedom & \checkmark \\
        consistency properties & (3/3) \\
      \end{tabular}
      & 0.20 & 12 / 44 & 11 / 30 & 8 / 23 & 7 / 20 \\
      \hline

      Dijkstra-Scholten &
      \begin{tabular}{ll}
        deadlock-freedom & \checkmark \\
      \end{tabular}
      & 0.16 & 13 / 53 & 11 / 31 & 10 / 37 & 9 / 31 \\
      \hline

      Bakery &
      \begin{tabular}{ll}
        deadlock-freedom & \checkmark \\
        mutual exclusion & \checkmark \\
      \end{tabular}
      & 0.18 & 10 / 29 & 10 / 25 & 8 / 23 & 7 / 20 \\
      \hline

      Burns &
      \begin{tabular}{ll}
        deadlock-freedom & \checkmark \\
        mutual exclusion & \checkmark \\
      \end{tabular}
      & 0.19 & 10 / 84 & 9 / 40 & 8 / 35 & 7 / 32 \\
      \hline

      Dijkstra &
      \begin{tabular}{ll}
        deadlock-freedom & \checkmark \\
        mutual exclusion & \checkmark \\
      \end{tabular}
      & 35.68 & 375 / 13164 & 106 / 4080 & 13 / 162 & 10 / 138 \\
      \hline

      Broadcast MutEx &
      \begin{tabular}{ll}
        deadlock-freedom & \checkmark \\
        mutual exclusion & \checkmark \\
      \end{tabular}
      & 0.15 & 9 / 23 & 9 / 19 & 8 / 19 & 7 / 16 \\
      \hline

      Preemptive (high) &
      \begin{tabular}{ll}
        deadlock-freedom & \checkmark \\
        mutual exclusion & $\times$ \\
      \end{tabular}
      & 0.18 & 41 / 293 & 17 / 67 & 3 / 3 & 6 / 8 \\
      \hline

      Preemptive &
      \begin{tabular}{ll}
        deadlock-freedom & \checkmark \\
        mutual exclusion & \checkmark \\
      \end{tabular}
      & 0.17 & 28 / 180 & 22 / 102 & 18 / 101 & 12 / 64 \\
      \hline

      Preemptive (uninitialized) &
      \begin{tabular}{ll}
        deadlock-freedom & \checkmark \\
        mutual exclusion & $\times$ \\
      \end{tabular}
      & 0.16 & 20 / 80 & 11 / 37 & 8 / 23 & 7 / 20 \\
      \hline

      Semaphore &
      \begin{tabular}{ll}
        deadlock-freedom & \checkmark \\
        mutual exclusion & \checkmark \\
      \end{tabular}
      & 0.15 & 14 / 41 & 9 / 31 & 10 / 36 & 8 / 31 \\
      \hline

      Szymanski &
      \begin{tabular}{ll}
        deadlock-freedom & n.a. \\
        mutual exclusion & n.a. \\
      \end{tabular}
      & 36.61 & n.a. & 495 / 39202 & 8 / 95 & 7 / 108 \\
      \hline

      Dijkstra (ring) &
      \begin{tabular}{ll}
        deadlock-freedom & n.a. \\
        mutual exclusion & n.a. \\
      \end{tabular}
      & 83.46 & 703 / 9287 & 1149 / 20455 & 20 / 244 & 20 / 346 \\
      \hline

      Dining Cryptographers &
      \begin{tabular}{ll}
        deadlock-freedom & $\times$ \\
        correctness & $\checkmark$ \\
      \end{tabular}
      & 2.60 & 250 / 4396 & 288 / 4145 & 10 / 78 & 9 / 87 \\
      \hline

      Dining Philosophers (global) &
      \begin{tabular}{ll}
        deadlock-freedom & $\checkmark$ \\
      \end{tabular}
      & 0.23 & 32 / 174 & 19 / 112 & 15 / 80 & 11 / 59 \\
      \hline

      Herman (linear) &
      \begin{tabular}{ll}
        deadlock-freedom & $\times$ \\
        no token loss & $\checkmark$ \\
      \end{tabular}
      & 0.17 & 19 / 70 & 14 / 42 & 10 / 33 & 9 / 27 \\
      \hline

      Herman (ring) &
      \begin{tabular}{ll}
        deadlock-freedom & $\checkmark$ \\
        no token loss & $\checkmark$ \\
      \end{tabular}
      & 0.17 & 19 / 71 & 14 / 42 & 10 / 33 & 9 / 27 \\
      \hline

      Israeli-Jalfon &
      \begin{tabular}{ll}
        deadlock-freedom & $\checkmark$ \\
        no token loss & $\checkmark$ \\
      \end{tabular}
      & 0.17 & 43 / 185 & 14 / 40 & 3 / 3 & 6 / 8 \\
      \hline

      Dining Philosophers (lefty) &
      \begin{tabular}{ll}
        deadlock-freedom & $\checkmark$ \\
      \end{tabular}
      & 0.16 & 37 / 187 & 27 / 149 & 12 / 64 & 11 / 54 \\
      \hline

      Dining Philosophers (lefty, rem.~forks) &
      \begin{tabular}{ll}
        deadlock-freedom & $\checkmark$ \\
      \end{tabular}
      & 0.17 & 37 / 243 & 21 / 118 & 14 / 112 & 11 / 67 \\
      \hline

      Lehmann-Rabin &
      \begin{tabular}{ll}
        deadlock-freedom & $\checkmark$ \\
      \end{tabular}
      & 0.17 & 39 / 452 & 23 / 208 & 11 / 68 & 11 / 83 \\
    \end{tabular}
  }
\end{center}

%% file: conclusions.tex
\section{Conclusions}
We have shown that the trap technique used
in~\cite{BensalemBNS09,ELMMN14,BFHH16} for the verification of single
systems can be extended to parameterized systems with sophisticated
communication structures, like pipelines, token rings and trees. Our
extension constructs a parameterized trap invariant, a formula of
\wss{\kappa} satisfied by the reachable global states of all instances
of the system. The core of the approach is a purely syntactic,
automatic derivation of the trap invariant from the interaction
formula describing the possible transitions of the system. When the
set of safe global states can also be expressed in \wss{\kappa}, which
is usually the case, we check using the Mona tool whether the trap
invariant implies the safety formula. The technique proves correctness
of systems that do not produce well-structured transition systems in
the sense of~\cite{AbdullaCJT96,FinkelS01}, and of systems with
broadcast communication, for which, to the best of our knowledge,
cut-off results have not been obtained yet.

Our experiments demonstrate that trap invariants can be very effective
in finding proofs of correctness (inductive invariants) of common
benchmark examples. In practice, the technique is very cheap, since it
avoids costly fixpoint computations. This suggests incorporating it
into other verifiers as a cheap preprocessing step.

\subsubsection{Acknowledgements.} \small{The work of the
second and fifth author has received funding from the European Research
Council (ERC) under the European Union's Horizon 2020 research and
innovation programme under grant agreement No 787367 (PaVeS).}

%% file: appendix.tex
\section{Proofs from Section \ref{sec:il-wss}}

\LemmaIlWss*
\begin{proof}
 By induction on the structure of $\phi$. Since
  $\mathit{Tr}(.)$ is homomorphic w.r.t the first-order connectives, 
  the only interesting case is when $\phi$ is $\succ_i(x)=y$.

\smallskip
\noindent ($\Rightarrow$)  If $(\Universe,\iota) \ilmodels
  \succ_i(x)=y$ then $\sigma_i(\iota(x))=\iota(y)$ and we distinguish two
  cases: \begin{compactitem}
  \item if $\iota(x)$ is not a maximal element of $\Universe$, then
    $\sigma_i(\iota(x))=\xi_i(\iota(x))$, hence $(\Universe,\iota) \wsmodels
    \neg\max(x) \wedge \succ_i(x)=y$,
  \item else, $\iota(x)$ is a maximal element of $\Universe$, we have
    $\xi_i(\iota(x))=\epsilon$, hence $(\Universe,\iota) \wsmodels \max(x)
    \wedge y=\wssroot$.
  \end{compactitem}
  In each case, we obtain $(\Universe,\iota) \wsmodels
  \mathit{Tr}(\succ_i(x)=y)$. 
  
\smallskip\noindent ($\Leftarrow$) Since $(\Universe,\iota)
  \wsmodels \mathit{Tr}(\succ_i(x)=y)$, we have two
  cases: \begin{compactitem}
  \item if $(\Universe,\iota) \wsmodels \neg\max(x) \wedge \succ_i(x)=y$
    then $\iota(x)$ is not a maximal element of $\Universe$ and
    $\xi_i(\iota(x))=\iota(y)$, thus $\sigma_i(\iota(x))=\iota(y)$ and
    $(\Universe,\iota) \wsmodels \succ_i(x)=y$.
  \item else $(\Universe,\iota) \wsmodels \max(x) \wedge
    y=\overline{\epsilon}$ and $\iota(x)$ is a maximal element of
    $\Universe$. In this case we have $\sigma_i(\iota(x))=\epsilon=\iota(y)$,
    hence $(\Universe,\iota) \ilmodels \succ_i(x)=y$. \qed
  \end{compactitem}
  \end{proof}

\section{Proofs from Section \ref{sec:partrap}}

\LemmaTrappred*
\proof{ Central to our analysis of $\anet^{\Universe}_{\asys} =
  (\places,\trans,\edges)$ is the following fact:
  \begin{fact}\label{fact:model-transition}
    For any transition $t \in \trans$ exists a clause
    \begin{equation*}
      \aclause(x_1, \ldots, x_\ell)= \begin{array}{c}
       \exists  ~.~ \varphi \wedge
       \bigwedge_{j=1}^\ell p_j(x_j) \wedge \bigwedge_{j=\ell+1}^{\ell+m}
       \forall x_j ~.~ \psi_j \rightarrow p_j(x_j)
     \end{array}
    \end{equation*}
    in $\interform$ and values $u_1,\ldots,u_\ell \in \Universe$ such that:
    \begin{equation*}
      \small{
        \begin{array}{l}
          \tuple{\Universe, \iota[x_{1} \leftarrow u_{1}, \ldots, x_{\ell} \leftarrow
          u_{\ell}]} \ilmodels \varphi,\\\\
            \pre{t} =
              \set{\tuple{s, u}\mid u \in \Psi_{j}
             \text{ and }s = \pre{p_{j}}\text{ for some }\ell + 1 \leq j\leq
             \ell + m}
             \cup\set{\tuple{\pre{p_{1}}, u_{1}}, \ldots, \tuple{
               \pre{p_{\ell}}, u_{\ell}}} \\\\
           \post{t} =
             \set{\tuple{s, u}\mid u \in \Psi_{j}
             \text{ and }s = \post{p_{j}}\text{ for some }\ell + 1 \leq j\leq
             \ell + m}
             \cup\set{\tuple{\post{p_{1}}, u_{1}}, \ldots, \tuple{
               \post{p_{\ell}}, u_{\ell}}}
       \end{array}
     }
   \end{equation*}
   where $\Psi_{j} \isdef \set{u\in U\mid \tuple{\Universe, \nu[x_{j}
         \leftarrow u], \emptyset} \ilmodels \psi_{j}}$ for $\ell + 1
   \leq j \leq \ell + m$.  Dually, for a fixed clause $\aclause$ in
   $\interform$ and a tuple of values $\overline{u} = \tuple{u_{1},
     \ldots, u_{\ell}}$ such that $\tuple{\Universe, \iota[x_{1} \leftarrow
       u_{1}, \ldots, x_{\ell} \leftarrow u_{\ell}]} \ilmodels
   \varphi$ there exists a unique transition $t \in \trans$ with
   $\pre{t}$ and $\post{t}$ as above, denoted $\tuple{\aclause,
     \overline{u}}$. 

 \end{fact}
 \proof{ By the definition of $\anet^\Universe_\asys$, for any $t \in \trans$
   there exists a clause $\aclause$ of $\interform$ of the form
   (\ref{eq:param-interform}) and a minimal model $\I = \tuple{\Universe,
     \iota}$ of $\aclause$.  Hence, there are values $u_{1} \in
   \iota(p_{1}), \ldots, u_{\ell} \in \iota(p_{\ell})$ such that
   $\tuple{\Universe, \iota[x_{1} \leftarrow u_{1}, \ldots, x_{\ell}
       \leftarrow u_{\ell}]} \ilmodels \varphi$. By the minimality of
   $\I$, $u_1, \ldots, u_\ell$ are the only elements of $\iota(p_1),
   \ldots, \iota(p_\ell)$, respectively. Moreover, for every $\ell+1
   \leq j \leq \ell+m$ and $u \in \Psi_j$, we have $u \in \iota(p_j)$
   and nothing else is in $\iota(p_j)$. By the definition of $\anet^\Universe_\asys$, we obtain
   $\pre{t}$ and $\post{t}$ as stated above.

   For the dual direction, we fix $\aclause$ in $\interform$ and $u_1,
   \ldots, u_\ell$ such that $\tuple{\Universe, \iota[x_{1} \leftarrow u_{1},
       \ldots, x_{\ell} \leftarrow u_{\ell}]} \ilmodels \varphi$. The
   structure $\overline{\I} = (\Universe,\overline{\iota})$, where: 
   \[\begin{array}{rcl}
   \overline{\iota}(x_j) & = & u_j \text{ and } \overline{\iota}(p_j) = \set{u_j}\text{, for all $1 \leq j \leq \ell$} \\
   \overline{\iota}(p_j) & = & \Psi_j \text{, for all $\ell+1 \leq j \leq \ell+m$}
   \end{array}\]
   is a minimal model of $\aclause$ and uniquely determines $t \in T$
   with the above properties. \qed}
 
 Next, we prove the following points: 
 \begin{compactenum}
 \item\label{item:preset} $P\cap\pre{\tuple{\aclause, \iota(x_{1}),
     \ldots, \iota(x_{\ell})}} \neq \emptyset \iff (\Universe, \iota)
   \wsmodels \intersectspre(\overline{\statevar{}}, x_{1}, \ldots,
   x_{\ell})$:

   \paragraph{\enquote{$\Rightarrow$}}
   If $P\cap\pre{\tuple{\aclause, \iota(x_{1}), \ldots,
       \iota(x_{\ell})}} \neq \emptyset$, then we have two
   cases: \begin{compactitem}
   \item if $\tuple{\pre{p_{i}}, \iota(x_{i})} \in P$, for some $1
     \leq i \leq \ell$, then by the definition of $P$, we obtain
     $\iota(x_{i}) \in \iota(X_{\pre{p_{i}}~})$ and therefore $(\Universe,
     \iota) \wsmodels X_{\pre{p_{i}}~}(x_{i})$,
   \item else $\tuple{\pre{p_{j}}, u}\in P$, for one $\ell + 1 \leq j
     \leq \ell + m$, leading to $u \in \Psi_j$. But then
     $(\Universe,\iota[x_j \leftarrow u]) \ilmodels \psi_j$, hence
     $(\Universe,\iota[x_j \leftarrow u]) \wsmodels \mathit{Tr}(\psi_j)$, by
     Lemma \ref{lemma:il-wss}. Since also $(\Universe,\iota[x_j \leftarrow
       u]) \wsmodels X_{\pre{p_{j}}~}(x_{j})$, by a similar argument
     as before, we obtain $(\Universe, \iota) \wsmodels \exists x_{j}~.~
     \mathit{Tr}(\psi_{j}) \wedge \statevar{\pre{p_{j}}~}(x_{j})$.
   \end{compactitem}

   \paragraph{\enquote{$\Leftarrow$}}
   If $(\Universe, \iota) \wsmodels
   \intersectspre(\overline{\statevar{}}, x_{1}, \ldots, x_{\ell})$
   then one of the following holds: \begin{compactitem}
   \item if $(\Universe, \iota) \wsmodels \statevar{\pre{p_{i}}~}(x_{i})$ for
     some $1 \leq i \leq \ell$, then, by the definition of $P$ and
     Fact \ref{fact:model-transition}, we obtain $\tuple{\pre{p_{i}},
       \nu(x_{i})} \in P\cap\pre{\tuple{\aclause, \iota(x_{1}),
         \ldots, \iota(x_{\ell})}}$.
   \item else $(\Universe,\iota) \wsmodels \exists x_j ~.~
     \mathit{Tr}(\psi_j) \wedge \statevar{\pre{p_j}~}(x_j)$ for some
     $\ell + 1 \leq j \leq \ell + m$, leading to the existence of a
     node $u_j \in \Universe$ such that $(\Universe, \iota[x_{j}\leftarrow u_{j}]) \wsmodels
   \mathit{Tr}(\psi_j) \wedge \statevar{\pre{p_j}~}(x_j)$ which, by
   Lemma \ref{lemma:il-wss}, yields $(\Universe, \iota[x_{j}\leftarrow
       c_{j}]) \ilmodels \psi_j \wedge
   \statevar{\pre{p_j}~}(x_j)$. This in turn leads to 
   $u_{j} \in \Psi_{j}$ and therefore $\tuple{\pre{p_{j}}, u_{j}} \in
   P\cap\pre{\tuple{\aclause, \iota(x_{1}), \ldots,
       \iota(x_{\ell})}}$. 
   \end{compactitem}
 \item\label{item:postset} $P\cap\post{\tuple{\aclause, \iota(x_{1}),
     \ldots, \iota(x_{\ell})}} \neq \emptyset \iff (\Universe, \iota)
   \wsmodels \intersectspost(\overline{\statevar{}}, x_{1}, \ldots,
   x_{\ell})$

   This case is proved using a similar argument as (\ref{item:preset})
   above.
\end{compactenum}
 Back to the main proof, we need to show that $P$ is a trap in
 $\anet^\Universe_\asys \iff (\Universe,\iota) \wsmodels
 \trappred(\overline{\statevar{}})$.

 \paragraph{\enquote{$\Rightarrow$}}
 Consider the case where $(\Universe, \iota) \not \wsmodels \trappred$.
 Hence, there is a clause $\aclause$ and nodes $u_{1}, \ldots,
 u_{\ell} \in \Universe$ such that $(\Universe, \iota[x_{1}\leftarrow u_{1}, \ldots,
   x_{\ell}\leftarrow u_{\ell}]) \wsmodels \mathit{Tr}(\varphi)$ thus,
 by Lemma \ref{lemma:il-wss}, $(\Universe, \iota[x_{1}\leftarrow u_{1},
   \ldots, x_{\ell} \leftarrow u_{\ell}]) \ilmodels \varphi$, and
 $(\Universe, \iota[x_{1} \leftarrow u_{1}, \ldots, x_{\ell}\leftarrow
   u_{\ell}]) \wsmodels \intersectspre$, whereas $(\Universe,
 \iota[x_{1}\leftarrow u_{1}, \ldots, x_{\ell} \leftarrow u_{\ell}])
 \not \wsmodels \intersectspost$. From \ref{item:preset} and
 \ref{item:postset}, we conclude that $P \cap \pre{\tuple{\aclause,
     \iota(x_{1}), \ldots, \iota(x_{\ell})}} \neq \emptyset$, whereas
 $P \cap \post{\tuple{\aclause, \iota(x_{1}), \ldots,
     \iota(x_{\ell})}} = \emptyset$. But then, $P$ is not a trap of
 $\anet^\Universe_\asys$, because $t \in \pre{P}$, whereas $t \not\in
 \post{P}$.

 \paragraph{\enquote{$\Leftarrow$}}
 Assume that $(\Universe, \iota) \wsmodels \trappred$ and let $t \in
 \post{P}$ be a transition. By Fact \ref{fact:model-transition}, there
 is a clause $\aclause$ and nodes $\overline{u}=\tuple{u_{1}, \ldots,
   u_{\ell}} \in \Universe$, such that $t = \tuple{\aclause,
   \overline{u}}$. Then, we have $(\Universe, \iota[x_{1} \leftarrow u_{1},
   \ldots, x_{\ell}\leftarrow u_{\ell}]) \ilmodels \varphi$, hence
 $(\Universe, \iota[x_{1} \leftarrow u_{1}, \ldots, x_{\ell}\leftarrow
   u_{\ell}]) \wsmodels \mathit{Tr}(\varphi)$, by Lemma
 \ref{lemma:il-wss}. Since $t \in \post{P}$, we have $P \cap \pre{t}
 \neq \emptyset$, thus $(\Universe, \iota[x_{1} \leftarrow u_{1}, \ldots,
   x_{\ell}\leftarrow u_{\ell}) \wsmodels \intersectspre$, by
   (\ref{item:preset}). Because $(\Universe, \iota) \wsmodels \trappred$, we
   also obtain that $(\Universe, \iota[x_{1} \leftarrow u_{1}, \ldots,
     x_{\ell}\leftarrow u_{\ell}) \wsmodels \intersectspost$, thus $P
     \cap \post{t} \neq \emptyset$, by (\ref{item:postset}). This
     leads to $t \in \pre{P}$, and since the choice of $t$ was
     arbitrary, we obtain $\post{P} \subseteq \pre{P}$, as
     required. 
\qed}

\TrapInvariant*
\proof{ Let $(\Universe,\iota)$ be a structure, such that $\iota$ interprets
  $\overline{\statevar{}}$ and $\overline{\statevar{}'}$ and let
  $\amark$ be a marking of $\anet^\Universe_\asys$. We define the sets: 
  \begin{equation*}
    \small{
      \begin{array}{l}
        \placeset_{\iota} \isdef \set{\tuple{s, u} \in \bigcup_{i = 1}^{K}
          \typeno{\states}{i} \times \Universe \mid u \in \iota(\statevar{s})
        }\text{ and }
        \placeset_{\iota}' \isdef \set{\tuple{s, u} \in \bigcup_{i = 1}^{K}
        \typeno{\states}{i} \times \Universe \mid u \in \iota(\statevar{s}')
        }
        \\
        \placeset_{\amark} \isdef \bigcup\limits_{s \in \bigcup_{i = 1}^{K}\typeno{\states}{i}}
        \set{s} \times \placeset_{\amark}^{s}\text{ where }\placeset_{\amark}^{s}
        \isdef \set{ u \in \Universe \mid \amark(\tuple{s, u}) = 1
        }\\
      \end{array}
    }
  \end{equation*}
  and note the following points (the proofs of which are easy checks
  left to the reader):
  \begin{enumerate}[(a)]
    \item\label{item:marking} for any marking $\amark$ of $\anet^{\Universe}_{\asys}$
      if $\iota(\statevar{s}) = \placeset_{\amark}^{s}$ then
      $(\Universe,\iota) \wsmodels \marking(\overline{\statevar{}})$,
    \item\label{item:intersection} $(\Universe, \iota) \wsmodels
      \intersection(\overline{\statevar{}},\overline{\statevar{}'})
      \iff \placeset_{\iota} \cap \placeset_{\iota}' \neq \emptyset$, 
      and
    \item\label{item:initially-marked} if $\amark_{0}$ is the initial marking of
      $\amarkednet^{\Universe}_{\asys}$, then
      \begin{equation*}
        \small{
          (\Universe,\iota) \wsmodels \initially(\overline{\statevar{}}) \iff \placeset{\iota}\cap
          \placeset_{\amark_{0}} \neq \emptyset.
        }
      \end{equation*}
  \end{enumerate}
  With (\ref{item:initially-marked}) and Lemma \ref{lemma:trappred} we
  conclude that $(\Universe, \iota) \wsmodels
  \trappred(\overline{\statevar{}'})
  \wedge\initially(\overline{\statevar{}'})$ if and only if
  $\placeset_{\iota}'$ is a IMT. Which yields using
  (\ref{item:intersection}) that $(\Universe,\iota) \wsmodels
  \trapconstraintpred$ if and only if $\placeset_{\iota}$ intersects
  all IMTs. Hence, if there is a marking $\amark \in
  \reach{\amarkednet^{\Universe}_{\asys}}$ such that $(\Universe,\iota[\statevar{s}
    \leftarrow \placeset_{\amark}^{s}]_{s \in \cup_{i =
      1}^{K}\typeno{\states}{i}}) \wsmodels \neg\varphi$,
  i.e.\ $\placeset_{\amark}$ \enquote{violates} $\varphi$, then by
  (\ref{item:marking}), we obtain:
  \begin{equation*}
     \small{
       (\Universe,\iota[\statevar{s} \leftarrow \placeset_{\amark}^{s}]_{s
       \in \cup_{i = 1}^{K}\typeno{\states}{i}} \wsmodels
       \marking(\overline{\statevar{}})\wedge
       \trapconstraintpred(\overline{\statevar{}})\wedge
       \neg\varphi(\overline{\statevar{}}).
       }
  \end{equation*}
  The proof follows by contraposition. \qed
}

\section{Proofs from Section \ref{sec:refine-trap-inv}}

\Cex* 
\proof{ Let $C = \set{\tuple{b,0}, \tuple{h,0}, \tuple{b,1},
    \tuple{w,1}, \tuple{f,2}, \tuple{e,2}}$ in the following. We shall
  try to build a nonempty trap $T$ that avoids every state in $C$. If
  such a trap can be found, the counterexample is shown to be spurious
  (unreachable). Below is the list of states allowed in $T$, indexed
  by component (using other states that the ones listed below would
  result in a trap that is satisfied by the counterexample $C$, which
  is exactly the opposite of what we want):

\[\begin{array}{c|c|c|c|c|c}
\mathsf{Fork}(0) & \mathsf{Philosopher}_{rl}(0) & \mathsf{Fork}(1) & 
\mathsf{Philosopher}_{lr}(1) & \mathsf{Fork}(2) & \mathsf{Philosopher}_{lr}(2) \\
\hline
\tuple{f,0} & \tuple{w,0}, \tuple{e,0} & \tuple{f,1} & \tuple{h,1}, \tuple{e,1} & \tuple{b,2} & \tuple{w,2}, \tuple{h,2}
\end{array}\]

Assume that $\tuple{f,0} \in T$. Then $T$ must contain $\tuple{b,0}$ or $\tuple{e,2}$
(constraint $gr(2) \wedge g(0)$). However neither is allowed, thus
$\tuple{f,0} \not\in T$. Assume that $\tuple{f,1} \in T$. Then $T$ must contain
$\tuple{b,1}$ or $\tuple{h,0}$ (constraint $gr(0) \wedge g(1)$), contradiction, thus
$\tuple{f,1} \not\in T$. Assume that $\tuple{b,2} \in T$. Then $T$ must contain
$\tuple{f,1}, \tuple{w,1}$ or $\tuple{f,2}$ (constraint $p(1) \wedge \ell(1) \wedge
\ell(2)$), contradiction, thus $\tuple{b,2} \not\in T$. Then $T$ contains
only philosopher states, except for $\tuple{h,0}$, $\tuple{w,1}$ and $\tuple{e,2}$. One can
prove that there is no such trap, for instance, for
$\mathsf{Philosopher}_{lr}(1)$ we have:
\[\begin{array}{rcl}
\tuple{h,1} \in T & \Rightarrow & \tuple{e,1} \in T \\
\tuple{e,1} \in T & \Rightarrow & \tuple{w,1} \in T
\end{array}\]
since $\tuple{f,1}, \tuple{b,1}, \tuple{f,2}, \tuple{b,2} \not\in
T$. Since $\tuple{w,1} \not\in T$, we obtain that $\tuple{h,1},
\tuple{e,1} \not\in T$. Then the only possibility is
$T=\emptyset$. \qed}

\section{Missing Material from Section \ref{sec:flow-invariants}}

\OneInvariant*

\begin{proof}
   A simple inductive argument allows to prove the invariant
   $\sum\limits_{s\in \flow}\amark(s) = 1$ 
   for every marking $\amark \in
   \reach{\amarkednet}$ while additionally maintaining that any transition $t$
   such that $\card{W \cap \pre{t}~} > 1$ cannot be fired. \qed
\end{proof}

\newcommand{\vecx}{\vec{x}}
In the following we give the missing formulae and auxiliaries from Section
\ref{sec:flow-invariants} for the case that the preset of a transition is
considered. The case of a postset is completely analogous. 
We let $\vecx$ denote a vector $\vecx := (x_1, \ldots, x_\ell)$.

\begin{equation*}
  \small{
    \begin{array}{l}
      \uniqueinitially(\overline{\statevar{}}) = \exists x~.~
      \bigvee\limits_{1\leq c\leq K}
        \left(
          \statevar{\typeno{\initstate}{c}}(x)\wedge
          \bigwedge\limits_{1\leq o\leq K}^{o\neq c}\neg
          \statevar{\typeno{\initstate}{o}}
        \right)
        \wedge\forall y~.~\left[
          \bigvee\limits_{1\leq c\leq K}\statevar{\typeno{\initstate}{c}}(y)
        \right]\rightarrow x = y\\
    \end{array}
  }
\end{equation*}
\begin{equation*}
  \small{
    \begin{array}{l}
      \begin{array}{r}
        \uniqueintersection(\overline{\statevar{}}, \overline{\statevar{}'}) =
        \exists x~.~
        \bigvee\limits_{q \in \cup_{1\leq c\leq K}\typeno{\states}{c}}
          \left(
            (\statevar{q}(x) \wedge \statevar{q}'(x))
            \wedge \bigwedge\limits_{p \in \cup_{1\leq c\leq K}\typeno{\states}{c}\setminus\set{q}}\neg
            (\statevar{p}(x) \wedge \statevar{p}'(x))
          \right)\\
          \wedge\forall y~.~\left[
            \bigvee\limits_{q \in \cup_{1\leq c\leq K}\typeno{\states}{c}}
            \statevar{q}(y) \wedge \statevar{q}'(y)
          \right]\rightarrow x = y
      \end{array}
    \end{array}
  }
\end{equation*}

\begin{equation*}
  \small{
    \begin{array}{l}
      \uniquepreex(\overline{\statevar{}}, \vecx) =
      \bigvee\limits_{1 \leq i \leq \ell}
        \left(
          \statevar{\pre{p_{i}}~}(x_{i})\wedge
          \bigwedge\limits^{p_{o}\neq p_{i}}_{1\leq o\leq \ell}\neg
            \statevar{\pre{p_{o}}~}(x_{o})
            \wedge\bigwedge\limits^{p_{o} = p_{i}}_{1\leq o\leq \ell}
            \statevar{\pre{p_{o}}~}(x_{o})\rightarrow x_{i} = x_{o}
      \right)\\
    \end{array}
  }
\end{equation*}
\begin{equation*}
  \small{
    \begin{array}{l}
      \disjointprebroadcast(\overline{\statevar{}}, \vec{x}) =
      \forall y~.~\bigwedge\limits_{\ell + 1\leq j\leq \ell + m}
      \left[
        \psi(\vec{x}, y) \rightarrow \neg
        \statevar{\pre{p_{j}}~}(y)
      \right]\\
    \end{array}
  }
\end{equation*}
\begin{equation*}
  \small{
    \begin{array}{l}
      \disjointpreex(\overline{\statevar{}}, \vecx) =
      \bigwedge\limits_{1 \leq i \leq \ell}\neg\statevar{\pre{p_{i}}~}(x_{i})\\
    \end{array}
  }
\end{equation*}
\begin{equation*}
  \small{
    \begin{array}{l}
      \begin{array}{r}
      \uniqueprebroadcast(\overline{\statevar{}}, \vecx) =
        \bigvee\limits_{\ell+1 \leq j \leq \ell+m}\exists x_{j}~.~
          \Big(
            \psi_{j}(\vecx, x_{j})\wedge
            \statevar{\pre{p_{j}}~}(x_{j})\hfill\\
          \wedge \forall y~.~ \left[\psi_{j}(\vecx, y)
          \wedge \statevar{\pre{p_{j}}~}(y) \rightarrow y = x_{j}\right]\\
          \wedge \bigwedge\limits_{\ell + 1 \leq o \leq \ell + m}^{o \neq j}
          \left[
            \psi_{o}(\vecx, y) \rightarrow \neg \statevar{
              \pre{p_{o}}~}(y)
          \right]
          \Big)
      \end{array}
    \end{array}
  }
\end{equation*}
\begin{equation*}
 \small{
 \uniquepre(\overline{\statevar{}}, \vecx)  =     
 \begin{array}{l}
      \left(
        \uniquepreex(\overline{\statevar{}}, \vecx) \wedge
        \disjointprebroadcast(\overline{\statevar{}}, \vecx)
      \right) \vee \\
      \left(
        \uniqueprebroadcast(\overline{\statevar{}}, \vecx)
        \wedge \disjointpreex(\overline{\statevar{}}, \vecx)
      \right)
\end{array}
}
\end{equation*}

\LemmaFlowPred*
\begin{proof}
  We begin by formulating some auxilliary statements. For this we fix a
  clause $\aclause(x_{1}, \ldots, x_{\ell})$ in $\interform$ and values $u_{1},
  \ldots, u_{\ell}$ such that $(\Universe, \iota[x_{1} \leftarrow u_{1}, \ldots,
  x_{\ell} \leftarrow u_{\ell}]) \ilmodels \varphi$ and by Fact
  \ref{fact:model-transition} we get a corresponding transition $t$. We
  separate $\pre{t}$ and $\post{t}$ into its existential and universal part,
  i.e.
  \begin{equation*}
    \small{
      \pre{t} = \pre{t^{\exists}} \cup \pre{t^{\forall}}
        \text{ with }\pre{t^{\exists}} = \set{\tuple{\pre{p_{1}}, c_{1}},
        \ldots, \tuple{\pre{p_{\ell}}, c_{\ell}}}
        \text{ and }\pre{t^{\forall}} = \set{\set{\pre{p_{j}}~}\times
        \Psi_{j}: \ell + 1 \leq j \leq \ell + m}.
    }
  \end{equation*}
  W.l.o.g. we assume that $\pre{t^{\exists}} \cap \pre{t^{\forall}}=\emptyset$
  and moreover $\card{\set{p_{j} : \ell + 1 \leq j \leq \ell + m}} = m$. This
  is achieved by two steps. Namely, if there are two broadcasts
  \begin{equation*}
    \small{
      \forall x_{j_{1}} ~.~ \varphi_{j_{1}}(x_{1}, \ldots, x_{\ell}, x_{j_{1}})
      \rightarrow p_{j_{1}}(x_{j_{1}})
    \text{ and }
      \forall x_{j_{2}} ~.~ \varphi_{j_{2}}(x_{1}, \ldots, x_{\ell}, x_{j_{2}})
      \rightarrow p_{j_{2}}(x_{j_{2}})
    }
  \end{equation*}
  such that $p_{j_{1}} = p_{j_{2}} = p$ then we replace these by one
  broadcast\footnote{Note that this is not a valid broadcast in the sense of
  clauses, however in the translation into \wss{\kappa}{} we can
  incorporate these changes.}
  \begin{equation*}
    \small{
      \forall x_{j} ~.~ (\varphi_{j_{1}}(x_{1}, \ldots, x_{\ell}, x_{j}) \vee
      \varphi_{j_{2}}(x_{1}, \ldots, x_{\ell}, x_{j})) \rightarrow
      p(x_{j}).
    }
  \end{equation*}
  Moreover, if there is $1 \leq i \leq \ell$ with an atom $p_{i}(x_{i})$ in
  $\aclause$ and a broadcast
  \begin{equation*}
    \small{
      \forall x_{j} ~.~ \varphi_{j} \rightarrow p_{j}(x_{j})
    }
  \end{equation*}
  such that $p_{i} = p_{j}$ then we replace the broadcast\footnote{This
  actually is a valid broadcast.} with
  \begin{equation*}
    \small{
      \forall x_{j} ~.~ (\varphi_{j} \wedge x_{j} \neq x_{i}) \rightarrow
      p_{j}(x_{j}).
    }
  \end{equation*}
  This allows that there is for every port precisely one broadcast term and
  no broadcast \enquote{shadows} an atom of a free variable in $\aclause$.

  Our proof relies on the following observations:
  \begin{enumerate}[(a)]
    \item\label{item:uniqueinitially} $\card{T_{\iota}\cap\set{
        \set{\typeno{\initstate}{k}} \times \Universe : 1 \leq k \leq K}} = 1 \iff
        (\Universe, \iota) \wsmodels \uniqueinitially$,
    \item\label{item:uniquepreex} $\card{T_{\iota}\cap\pre{t^{\exists}}~} = 1
      \iff (\Universe, \iota[x_{1}
      \leftarrow u_{1}, \ldots, x_{\ell} \leftarrow u_{\ell}]) \wsmodels
      \uniquepreex$,
    \item\label{item:uniqueprebroadcast} $\card{T_{\iota}\cap
      \pre{t^{\forall}}~} = 1 \iff (\Universe, \iota[x_{1}
      \leftarrow u_{1}, \ldots, x_{\ell} \leftarrow u_{\ell}]) \wsmodels
      \uniqueprebroadcast$,
    \item\label{item:disjointpreex} $\card{T_{\iota}\cap\pre{t^{\exists}}~} = 0
      \iff (\Universe, \iota[x_{1}
      \leftarrow u_{1}, \ldots, x_{\ell} \leftarrow u_{\ell}]) \wsmodels
      \disjointpreex$ and
    \item\label{item:disjointprebroadcast} $\card{T_{\iota}\cap
      \pre{t^{\forall}}~} = 0 \iff (\Universe, \iota[x_{1}
      \leftarrow u_{1}, \ldots, x_{\ell} \leftarrow u_{\ell}]) \wsmodels
      \disjointprebroadcast$.
  \end{enumerate}
  Since the line of reasoning can be easily adapted for the different cases we
  restrict our arguments to (\ref{item:uniqueprebroadcast}) since it is the
  most elaborate formula.

  \paragraph{ad (\ref{item:uniqueprebroadcast}) \enquote{$\Rightarrow$}:}
  Fix $j \in \set{\ell + 1, \ldots, \ell + m}$ such that $q = \pre{p_{j}}$, and
  $u_{j} \in \Psi_{j}$ with $\set{\tuple{q, u_{j}}} = T_{\iota}\cap
  \pre{t^{\forall}}$. Hence, by definition of $T_{\iota}$, $u_{j} \in
  \iota(\statevar{q})$ and therefore $(\Universe, \iota[x_{1} \leftarrow u_{1},
  \ldots, x_{\ell} \leftarrow u_{\ell}, x_{j} \leftarrow u_{j}]) \wsmodels
  \statevar{q}(x_{j})$, while, by definition of $\Psi_{j}$ and Lemma
  \ref{lemma:il-wss},
  $(\Universe, \iota[x_{1} \leftarrow u_{1}, \ldots, x_{\ell} \leftarrow
  u_{\ell}, x_{j} \leftarrow u_{j}]) \wsmodels \mathit{Tr}(\psi_{j})$.
  For any fixed $u_{y} \in \cup_{\ell + 1 \leq k \leq \ell + m}
  \Psi_{k}$ we distinguish different cases:

  If $u_{y} = u_{j}$, we immediately have
  $(\Universe, \iota[x_{1} \leftarrow u_{1}, \ldots, x_{\ell} \leftarrow
  u_{\ell}, x_{j} \leftarrow u_{j}, y \leftarrow u_{y}]) \wsmodels
  \left[\psi_{j}(x_{1}, \ldots, x_{\ell}, y) \wedge \statevar{\pre{p_{j}}~}(y)
  \rightarrow y = x_{j}\right]$ and, by pairwise disjointness of all
  $\Psi_{n}$ for $\ell + 1 \leq n \leq \ell + m$, we get
  $$(\Universe, \iota[x_{1} \leftarrow u_{1}, \ldots, x_{\ell} \leftarrow
  u_{\ell}, x_{j} \leftarrow u_{j}, y \leftarrow u_{y}]) \wsmodels
  \bigwedge\limits_{\ell + 1 \leq o \leq \ell + m}^{o \neq j} \left[
  \psi_{o}(x_{1}, \ldots, x_{\ell}, y) \rightarrow \neg
  \statevar{\pre{p_{o}}~}(y)\right]$$ since
  $(\Universe, \iota[x_{1} \leftarrow u_{1}, \ldots, x_{\ell} \leftarrow
  u_{\ell}, x_{j} \leftarrow u_{j}, y \leftarrow u_{y}]) \not\wsmodels
  \psi_{o}(x_{1}, \ldots, x_{\ell}, y)$ for any $o \neq j$.

  If on the other hand, $u_{y} \in \Psi_{k}$ for $k \neq j$ then we have by
  definiton of $\Psi_{k}$ that $(\Universe, \iota[x_{1} \leftarrow
  u_{1}, \ldots, x_{\ell} \leftarrow u_{\ell}, x_{j} \leftarrow u_{j}, y
  \leftarrow u_{y}]) \wsmodels \mathit{Tr}(\psi_{k})$,
  however by pairwise disjointness of all $\Psi_{n}$ for $\ell + 1 \leq n \leq
  \ell + m$ we have $(\Universe, \iota[x_{1} \leftarrow u_{1}, \ldots, x_{\ell}
  \leftarrow u_{\ell}, x_{j} \leftarrow u_{j}, y \leftarrow u_{y}]) \not\wsmodels
  \mathit{Tr}(\psi_{n})$ for any
  $\ell + 1 \leq n \leq \ell + m$ different from $k$ and
  $(\Universe, \iota[x_{1} \leftarrow u_{1}, \ldots, x_{\ell} \leftarrow
  u_{\ell}, x_{j} \leftarrow u_{j}, y \leftarrow u_{y}]) \not\wsmodels
  \statevar{\pre{p_{k}}~}(y)$ because otherwise $\tuple{\pre{p_k}, u_{y}} \in
  T_{\iota}\cap\pre{t^{\exists}}$. This gives in combination
  $(\Universe, \iota[x_{1} \leftarrow u_{1}, \ldots, x_{\ell} \leftarrow
  u_{\ell}, x_{j} \leftarrow u_{j}, y \leftarrow u_{y}]) \wsmodels
  \bigwedge_{\ell + 1 \leq n \leq \ell + m} \psi_{n} \rightarrow \neg
  \statevar{\pre{p_{n}}~}(y)$ which already implies
  $$\begin{array}{l}
  (\Universe, \iota[x_{1} \leftarrow u_{1}, \ldots, x_{\ell} \leftarrow u_{\ell},
  x_{j} \leftarrow u_{j}, y \leftarrow u_{y}]) \wsmodels \\
  \hspace{1cm}
  \left[\psi_{j}(x_{1}, \ldots, x_{\ell}, y) \wedge \statevar{\pre{p_{j}}~}(y)
  \rightarrow y = x_{j}\right]\wedge
  \bigwedge\limits_{\ell + 1 \leq o \leq \ell + m}^{o \neq j} \left[
  \psi_{o}(x_{1}, \ldots, x_{\ell}, y) \rightarrow \neg
  \statevar{\pre{p_{o}}~}(y)\right]
  \end{array}$$

  Thirdly, consider the case that $u_{y} \in \Psi_{j}$ but $u_{y} \neq u_{j}$.
  Again by pairwise disjointness of all $\Psi_{n}$ for $\ell + 1 \leq n \leq
  \ell + m$ we have $(\Universe, \iota[x_{1} \leftarrow u_{1}, \ldots, x_{\ell}
  \leftarrow u_{\ell}, x_{j} \leftarrow u_{j}, y \leftarrow u_{y}])
  \not \wsmodels \mathit{Tr}(\psi_{n})$ for any $n$ different from $j$.
  However, since $\tuple{q, c_{y}} \in \pre{t^{\forall}}$ it cannot be in
  $T_{\iota}$ since it is in $\pre{t^{\exists}}$ which gives $u_{y} \not \in
  \iota(\statevar{\pre{p_{j}}~})$ and therefore
  $(\Universe, \iota[x_{1} \leftarrow u_{1}, \ldots,
  x_{\ell} \leftarrow u_{\ell}, x_{j} \leftarrow u_{j}, y \leftarrow u_{y}])
  \wsmodels \neg \statevar{q}(y)$.

  Concludingly, we may use $u_{j}$ as the witness for the existential
  quantification and get
  $(\Universe, \iota[x_{1} \leftarrow u_{1}, \ldots, x_{\ell} \leftarrow
  u_{\ell}]) \wsmodels \uniqueprebroadcast$.

  \paragraph{ad (\ref{item:uniqueprebroadcast}) \enquote{$\Leftarrow$}:}
  Consider the case that $T_{\iota}\cap \pre{t^{\forall}} = \emptyset$, then
  for all $u_{j} \in \Psi_{j}$ holds
  $(\Universe, \iota[x_{1} \leftarrow u_{1}, \ldots, x_{\ell} \leftarrow u_{\ell},
  x_{j} \leftarrow u_{j}]) \not \wsmodels \statevar{\pre{p_{j}}~}(x_{j})$ since
  $u_{j} \not \in \iota(\statevar{\pre{p_{j}}~})$
  for all $\ell + 1 \leq j \leq \ell + m$. Moreover, $(\Universe, \iota[x_{1}
  \leftarrow u_{1}, \ldots, x_{\ell} \leftarrow u_{\ell}, x_{j} \leftarrow
  u_{j}]) \wsmodels \neg \mathit{Tr}(\psi_{n})$ for all $\ell + 1 \leq n
  \leq \ell + m$ different from $j$ for all $\ell + 1 \leq j \leq \ell + m$ (
  again by pairwise disjointness). This specifically yields
  $(\Universe, \iota[x_{1} \leftarrow u_{1}, \ldots, x_{\ell} \leftarrow u_{\ell}],
  \iota) \wsmodels \neg
  \bigvee\limits_{\ell+1 \leq j \leq \ell+m}\exists x_{j}~.~ \psi_{j}
  \wedge \statevar{\pre{p_{j}}~}(x_{j})$.

  If on the other hand $\card{T_{\iota}\cap \pre{t^{\forall}}~} > 1$
  then we can fix an arbitrary $\tuple{q_{1}, u_{1}} \in T_{\iota}\cap
  \pre{t^{\forall}}$ with $u_{1} \in \Psi_{j_{1}}$ for $\ell + 1 \leq j_{1}
  \leq \ell + m$. Hence, $(\Universe, \iota[x_{1} \leftarrow u_{1}, \ldots,
  x_{\ell} \leftarrow u_{\ell}, x_{j_{1}} \leftarrow u_{1}]) \wsmodels
  \mathit{Tr}(\psi_{j_{1}}) \wedge \statevar{q_{1}}(x_{j_{1}})$. However, we
  always have a different $\tuple{q_{2}, u_{2}} \in T_{\iota}\cap \pre{t^{\forall}}$
  such that $u_{2} \in \Psi_{j_{2}}$ with $\ell + 1 \leq j_{2} \leq \ell + m$.
  Then, if $j_{1} = j_{2} = j$ we get $q_{1} = q_{2} = \pre{p_{j}}$ and
  therefore $(\Universe, \iota[x_{1} \leftarrow u_{1}, \ldots, x_{\ell} \leftarrow
  u_{\ell}, y \leftarrow u_{2}]) \wsmodels \mathit{Tr}(\psi_{j}(x_{1},
  \ldots, x_{\ell}, y)) \wedge \statevar{\pre{p_{j}}~}(y)$. Since $u_{1} \neq
  u_{2}$ we have $(\Universe, \iota[x_{1} \leftarrow u_{1}, \ldots, x_{\ell}
  \leftarrow u_{\ell}, x_{j} \leftarrow u_{1}]) \not \wsmodels
  \forall y~.~ \left[\psi_{j}(x_{1}, \ldots, x_{\ell}, y) \wedge
  \statevar{\pre{p_{j}}~}(y) \rightarrow y = x_{j}\right]$.

  If on the other hand $j_{1} \neq j_{2}$ then
  \begin{equation*}
    \small{
      (\Universe, \iota[x_{1} \leftarrow u_{1}, \ldots, x_{\ell} \leftarrow
      u_{\ell}, x_{j_{1}} \leftarrow u_{1}, y \leftarrow u_{2}]) \not \wsmodels
      \left[ \psi_{j_{2}}(x_{1}, \ldots, x_{\ell}, y)
      \rightarrow \neg \statevar{\pre{p_{j_{2}}}~}(y)\right]
    }
  \end{equation*}
  and consequently
  \begin{equation*}
    \small{
      (\Universe, \iota[x_{1} \leftarrow u_{1}, \ldots, x_{\ell} \leftarrow
      u_{\ell}, x_{j_{1}} \leftarrow u_{1}]) \not \wsmodels
      \bigwedge\limits_{\ell + 1 \leq o \leq \ell + m}^{o \neq j_{1}}
      \forall y ~.~ \left[ \psi_{o}(x_{1}, \ldots, x_{\ell}, y)
      \rightarrow \neg \statevar{\pre{p_{o}}~}(y)\right].
    }
  \end{equation*}

  However, for any value $u \in \Universe$ such that $(\Universe, \nu[x_{1} \leftarrow
  u_{1}, \ldots, x_{\ell} \leftarrow u_{\ell}, x_{j} \leftarrow u])
  \wsmodels \psi_{j}(x_{1}, \ldots, x_{\ell}, x_{j})\wedge
  \statevar{\pre{p_{j}}~}(x_{j})$ for one $\ell + 1 \leq j \leq \ell + m$
  it is easy to see from the definitions of $\pre{t^{\forall}}$ and $T_{\iota}$
  that $\tuple{\pre{p_{j}~}, c} \in \pre{t^{\forall}} \cap T_{\iota}$.
  This concludes then that $\tuple{\Universe, \nu[x_{1} \leftarrow
  c_{1}, \ldots, x_{\ell} \leftarrow c_{\ell}, x_{j} \leftarrow c_{1}], \iota}
  \not \wsmodels \uniqueprebroadcast$ because any choice for the existential
  quantification of $x$ does not allow for all values chosen for $y$ to satisfy
  the formula as demonstrated above.

  In consequence, the statements
  (\ref{item:uniqueinitially}) - (\ref{item:disjointprebroadcast}) show that
  the given formulae model the properties laid out in
  Lemma \ref{lemma:one-invariant} and therefore the lemma follows.
  \null\hfill\qed
\end{proof}

\FlowInvariant* \proof{ 
  In the light of Theorem
  \ref{thm:trap-invariant}, it is sufficient to prove the theorem for
  the formula:
  \begin{equation}
    \exists \overline{\statevar{}}~.~\marking(\overline{\statevar{}})\wedge
    \flowinvariant(\overline{\statevar{}})\wedge
    \neg\varphi(\overline{\statevar{}})
  \end{equation}
  The proof goes along the lines of the proof of Theorem
  \ref{thm:trap-invariant}, with the additional observation: 
  \[(\Universe, \iota) \wsmodels
      \uniqueintersection(\overline{\statevar{}},\overline{\statevar{}'})
      \iff \card{\placeset_{\iota} \cap \placeset_{\iota}'} = 1\]
  \qed}